\documentclass[aps,prb,twocolumn]{revtex4}

\usepackage{hyperref}
\usepackage{graphicx}
\usepackage[utf8]{inputenc}
\DeclareGraphicsExtensions{.png,.jpg,.eps,.pdf}
\usepackage{amsmath}
\usepackage{comment}
\usepackage{bm}
\usepackage{dsfont}
\usepackage{amsfonts}

\usepackage{xcolor}

\begin{document}

\title{Theory of triangulene two-dimensional crystals}

\author{
R. Ortiz$^{1,2}$ G. Catarina$^{1,3,4}$, J. Fern\'{a}ndez-Rossier$^{3}$\footnote{On permanent leave from Departamento de F\'{i}sica Aplicada, Universidad de Alicante, 03690 San Vicente del Raspeig, Spain}$^,$\footnote{joaquin.fernandez-rossier@inl.int}
}
\affiliation{
$^1$Departamento de F\'{i}sica Aplicada, Universidad de Alicante, 03690 San Vicente del Raspeig, Spain
\\
$^2$Donostia International Physics Center (DIPC), Paseo Manuel de Lardizabal 4, 20018 Donostia-San Sebasti\'{a}n, Spain
\\
$^3$Theory of Quantum Nanostructures Group, International Iberian Nanotechnology Laboratory (INL), Av. Mestre Jos\'{e} Veiga, 4715-330 Braga, Portugal
\\
$^4$Centro de F\'{i}sica das Universidades do Minho e do Porto, Universidade do Minho, Campus de Gualtar, 4710-057 Braga, Portugal
}

\date{\today}

%%%%%%%%%%%%%%%%%%%%%%%%%%%%%%%%%%%%%%%%%%%%%%%%%%%%%%%%%%%%%%%%%%%%%%%%%%%%%

\begin{abstract}

Equilateral triangle-shaped graphene nanoislands with a lateral dimension of $n$ benzene rings are known as $[n]$triangulenes.
Individual $[n]$triangulenes are open-shell molecules, with single-particle electronic spectra that host $n-1$ half-filled zero modes and a many-body ground state with spin $S=(n-1)/2$.
The on-surface synthesis of triangulenes has been demonstrated for $n=3,4,5,7$ and the observation of a Haldane symmetry-protected topological phase has been reported in chains of $[3]$triangulenes.
Here, we provide a unified theory for the electronic properties of a family of two-dimensional honeycomb lattices whose unit cell contains a pair of triangulenes with dimensions $n_a,n_b$. 
Combining density functional theory and tight-binding calculations, we find a wealth of half-filled narrow bands, including a graphene-like spectrum (for $n_a=n_b=2$), spin-1 Dirac electrons (for $n_a=2,n_b=3$), $p_{x,y}$-orbital physics (for $n_a=n_b=3$), as well as a gapped system with flat valence and conduction bands (for $n_a=n_b=4$).
All these results are rationalized with a class of effective Hamiltonians acting on the subspace of the zero-energy states that generalize the graphene honeycomb model to the case of fermions with an internal pseudospin degree of freedom with $C_3$ symmetry.

\end{abstract}

\maketitle

%%%%%%%%%%%%%%%%%%%%%%%%%%%%%%%%%%%%%%%%%%%%%%%%%%%%%%%%%%%%%%%%%%%%%%%%%%%%%

\section{Introduction}

The quest for new states of matter that do not occur naturally in conventional materials fuels the study of artificial quantum lattices in a variety of platforms, including cold atoms\cite{Hart2015,Mazurenko2017}, trapped ions\cite{blatt2012}, quantum dots\cite{Hensgens2017,Mortemousque2021}, dopants in silicon\cite{Salfi2016}, functionalized graphene bilayer\cite{GarciaMartinez2019}, moir\'{e} heterostructures\cite{kennes2021} and adatoms\cite{Yang2017,khajetoorians2019,Yang2021}.
These systems may work as quantum simulators of both spin and fermionic Hamiltonians, such as the Hubbard model, and are expected to host strongly correlated electronic phases.

Here, we explore the properties of two-dimensional (2D) artificial crystals with triangulenes as building blocks.
Triangulenes are open-shell polycyclic aromatic hydrocarbons.
Their low-energy degrees of freedom are provided by electrons that partially occupy zero-energy modes inside a large gap of strongly covalent molecular states formed by $\pi$ atomic orbitals.
In standard conditions, open-shell molecules are very reactive and therefore not suitable for manipulation.
However, thanks to the advances in on-surface chemical synthesis\cite{cai2010,Ruffieux16,su2020}, it has been possible to go around this problem, so that the controlled fabrication of triangulenes of various sizes\cite{pavlivcek2017,mishra2019b,su2019,Mishra2021b}, as well as triangulene dimers\cite{Mishra2020}, chains\cite{Mishra2021}, rings\cite{Mishra2021,hieulle2021} and other structures\cite{Cheng2022}, has been recently demonstrated.  
On the other hand, the on-surface synthesis of large-area carbon-based crystals has also been established\cite{steiner2017,moreno2018,telychko2021}.  
Therefore, the synthesis of 2D triangulene crystals seems within reach using state-of-the-art techniques, motivating the present work.

\begin{figure*}
\includegraphics[width=0.9\linewidth]{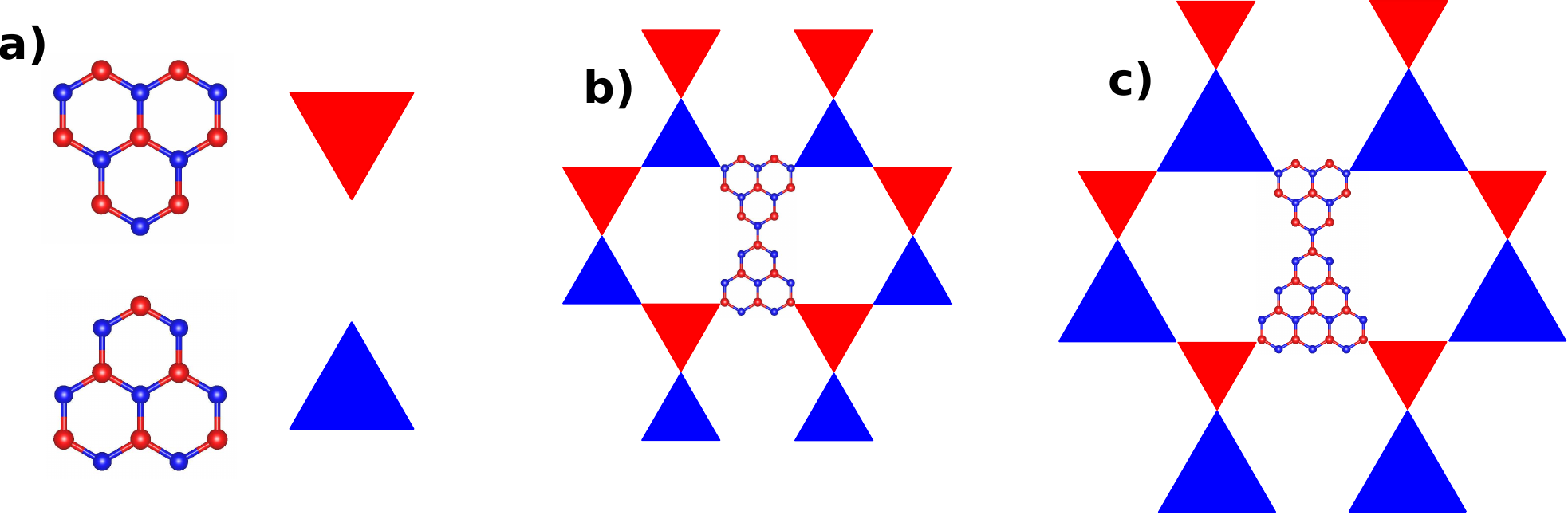}
\caption{
(a) Two variants of $[2]$triangulene, with $A$ ($B$) sublattice-type sites colored in red (blue).
The top (bottom) triangulene has excess of red (blue) sites.
(b) Centrosymmetric triangulene honeycomb lattice with a unit cell formed by a pair of $[2]$triangulenes with opposite sublattice imbalance.
(c) Non-centrosymmetric triangulene 2D crystal with a unit cell formed by an $A$-type $[2]$triangulene and a $B$-type $[3]$triangulene.
}
\label{fig:1}
\end{figure*}

Individual $[n]$triangulenes have a ground state with spin $S=(n-1)/2$, on account of their strong intramolecular exchange\cite{ovchinnikov78,JFR07}. 
The small spin-orbit coupling of carbon makes their magnetic anisotropy negligible\cite{Lado2014prl,Mishra2021}. 
The intermolecular exchange coupling in $[3]$triangulene dimers has been determined to be antiferromagnetic\cite{Mishra2020,Mishra2021,jacob2021,hieulle2021}. 
Recently reported\cite{Mishra2021} experimental results show that chains and rings of $[3]$triangulenes display the key features of the Haldane phase for antiferromagnetically coupled $S=1$ spins, namely a gap in the excitation spectrum and the emergence of fractional spin-1/2 edge states. 
Here, we study $[n_a,n_b]$triangulene 2D crystals, i.e., honeycomb lattices with a unit cell made of two triangulenes with dimensions $n_a,n_b$ (see Fig.~\ref{fig:1} for the cases $n_a=n_b=2$ and $n_a=2,n_b=3$).

The rest of this paper is organized as follows.
In section II, we review the electronic properties of individual $[n]$triangulenes.
We argue that, based on a nearest-neighbor tight-binding (TB) approximation, $[n_a,n_b]$triangulene 2D crystals should have $n_a+n_b-2$ {\em flat} bands at zero energy.
In section III, we calculate the energy spectrum using density functional theory (DFT) and we find $n_a+n_b-2$ {\em narrow} bands, in disagreement with the expectations based on the nearest-neighbor TB model.
We show that adding a third neighbor hopping to the TB model accounts for the DFT results and is essential to understand the weakly dispersive bands.
In section IV, we build a minimal TB model that includes only the zero-energy modes of the triangulenes and the effect of third neighbor hopping, compare it to both DFT and full TB models, and show that it accounts for the energy bands so obtained.
These include a graphene-like spectrum for $n_a=n_b=2$, spin-1 Dirac electrons for $n_a=2,n_b=3$, $p_{x,y}$-orbital honeycomb physics for $n_a=n_b=3$, and a gapped system with flat valence and conduction bands for $n_a=n_b=4$.
In section V, we address the effect of interactions and the spin physics in triangulene 2D crystals.
In section VI, we wrap up and present our conclusions. 

%%%%%%%%%%%%%%%%%%%%%%%%%%%%%%%%%%%%%%%%%%%%%%%%%%%%%%%%%%%%%%%%%%%%%%%%%%%%%

\section{Electronic properties of $[n]$triangulenes}

\subsection{Spectrum and symmetries}

\begin{figure*}
\includegraphics[width=0.9\linewidth]{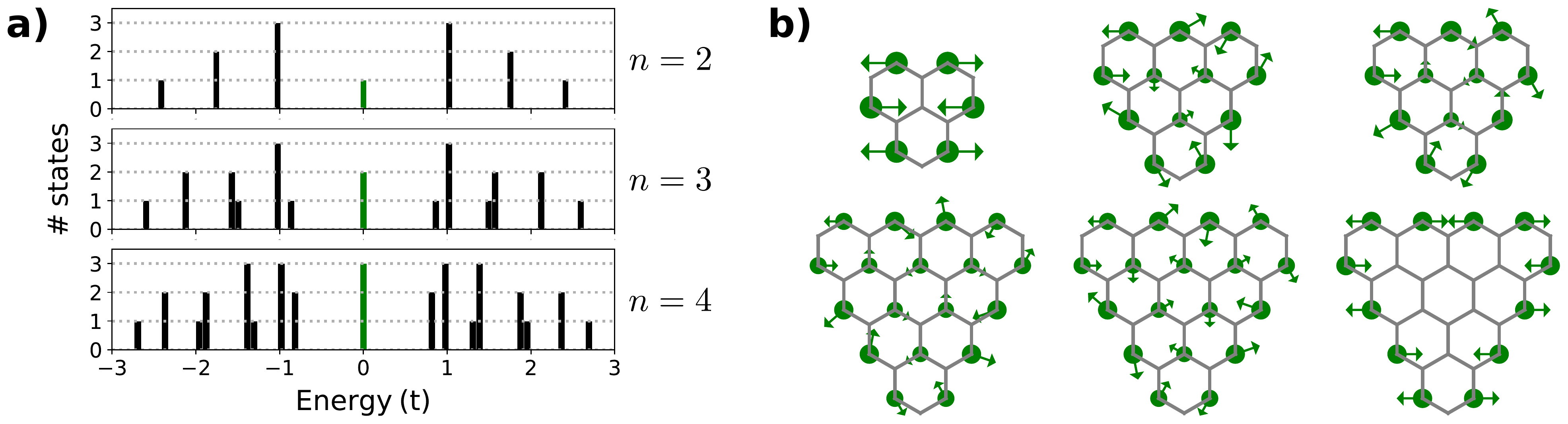}
\caption{
(a) Number of states vs energy of $[n]$triangulene, for $n=2,3,4$, calculated with a nearest-neighbor TB model, featuring $n-1$ zero-energy modes.
(b) Representation of the zero mode wave functions: the circle size scales with the modulus squared, the arrows represent the phases.
For $n=3,4$, the zero modes are chosen as eigenvectors of the $C_3$ symmetry operator (see Eq.~\eqref{eq:C3basis} and Table~\ref{tab:1}). 
The zero modes vanish at the minority sublattice, including the corner sites that link triangulenes to their neighbors.
}
\label{fig:2}
\end{figure*}

In this section, we briefly review the electronic properties of individual $[n]$triangulenes.
We first consider their single-particle properties, as described with the TB Hamiltonian, considering one orbital per site and nearest-neighbor hopping, $t$.
Because of the bipartite character of the lattice, the energy spectrum must be composed of two types of states: finite-energy states, with electron-hole symmetry, and at least $|N_A-N_B|$ zero modes, where $N_{A/B}$ is the number of sites of the $A/B$ sublattice. 
In the case of $[n]$triangulenes (see Fig.~\ref{fig:2}a for $n=2,3,4$), $|N_A-N_B| = n-1$ and we find $n-1$ zero modes.
We can build triangulenes with excess of either $A$ or $B$ sites (Fig.~\ref{fig:1}a).
We denote their zero mode wave functions by $|a\rangle$ and $|b\rangle$, respectively.

The zero modes are separated from the finite-energy states by a large energy splitting, as shown in Fig.~\ref{fig:2}a.
At half filling, relevant for polycyclic aromatic hydrocarbons at charge neutrality, the zero modes are half-filled.
Therefore, the low-energy physics of triangulene crystals are expected to be dominated by the electrons occupying the zero-energy states.

We now focus on the wave functions of the zero modes. 
As we show in Fig.~\ref{fig:2}b, the zero modes are sublattice-polarized, i.e., they have a non-zero weight exclusively in the {\em majority} sublattice\cite{JFR07,Ortiz19}. 
Because of the degeneracy of the zero modes for $n \geq 3$, their representation is not unique.
It is extremely useful to choose $|a\rangle$ and $|b\rangle$ as eigenstates of the $R_{2\pi/3}$ counterclockwise rotation operator:
\begin{equation}
R_{2\pi/3}|a\rangle= e^{i\omega_a}|a\rangle, \quad
R_{2\pi/3}|b\rangle= e^{i\omega_b}|b\rangle,
\label{eq:C3basis}
\end{equation}
where $e^{i3\omega_{a,b}}=1$.
We thus have three possible values for the exponents: $\omega_{a,b}=0,\pm 2\pi/3$.

\subsection{Interactions and magnetic properties}

The $n-1$ zero modes of the $[n]$triangulenes host $n-1$ electrons. 
Therefore, $[n]$triangulenes have an open-shell structure. 
Calculations carried out with DFT\cite{JFR07,wang09}, quantum chemistry\cite{Ortiz19} and model Hamiltonians\cite{JFR07,guclu09,Ortiz19} consistently predict that the many-body ground state maximizes the spin of the electrons in the zero modes, so that $2S=n-1$.  
When modeled with the Hubbard model, the spin of the ground state can be anticipated using Lieb's theorem\cite{Lieb89}, which states that $2S=|N_A-N_B|$. 
For $[n]$triangulenes, $|N_A-N_B|=n-1$, so that $2S=n-1$.  
The underlying mechanism for the ferromagnetic intra-triangulene exchange is a molecular version of the atomic Hund's coupling.

\subsection{Intermolecular hybridization of zero modes}

Let us now consider the unit cell of a $[n_a,n_b]$triangulene crystal, i.e., a dimer made of two triangulenes of $A$ and $B$ types.
Separately, these triangulenes would have $n_{a/b}-1$ zero modes each. 
When coupled together, Sutherland's theorem\cite{Sutherland1986} warrants a minimal number of $|n_a-n_b|$ zero modes per unit cell, and the same number of zero-energy ($E=0$) flat bands for the corresponding 2D crystal.
Specifically, for $n_a=n_b$ the theorem does {\em not} ensure the existence of $E=0$ flat bands.
However, the binding sites of the unit cell in the triangulene 2D lattice belong to the minority sublattice, whereas the zero modes are hosted in the majority sublattice (Fig.~\ref{fig:2}b).
Therefore, {\em nearest neighbor hopping does not hybridize zero modes of adjacent triangulenes}.
As a consequence, $t$ does not lift the zero mode degeneracy in a $[n_a,n_b]$triangulene crystal, so that a nearest neighbor TB model predicts $n_a+n_b-2$ flat bands at zero energy.
We anticipate that intermolecular hybridization of zero modes in triangulene 2D crystals will be governed by the small third neighbor hopping, leading to weakly dispersive half-filled energy bands.

%%%%%%%%%%%%%%%%%%%%%%%%%%%%%%%%%%%%%%%%%%%%%%%%%%%%%%%%%%%%%%%%%%%%%%%%%%%%%

\section{Non-magnetic energy bands: DFT vs TB}

\begin{figure*}
\includegraphics[width=0.9\linewidth]{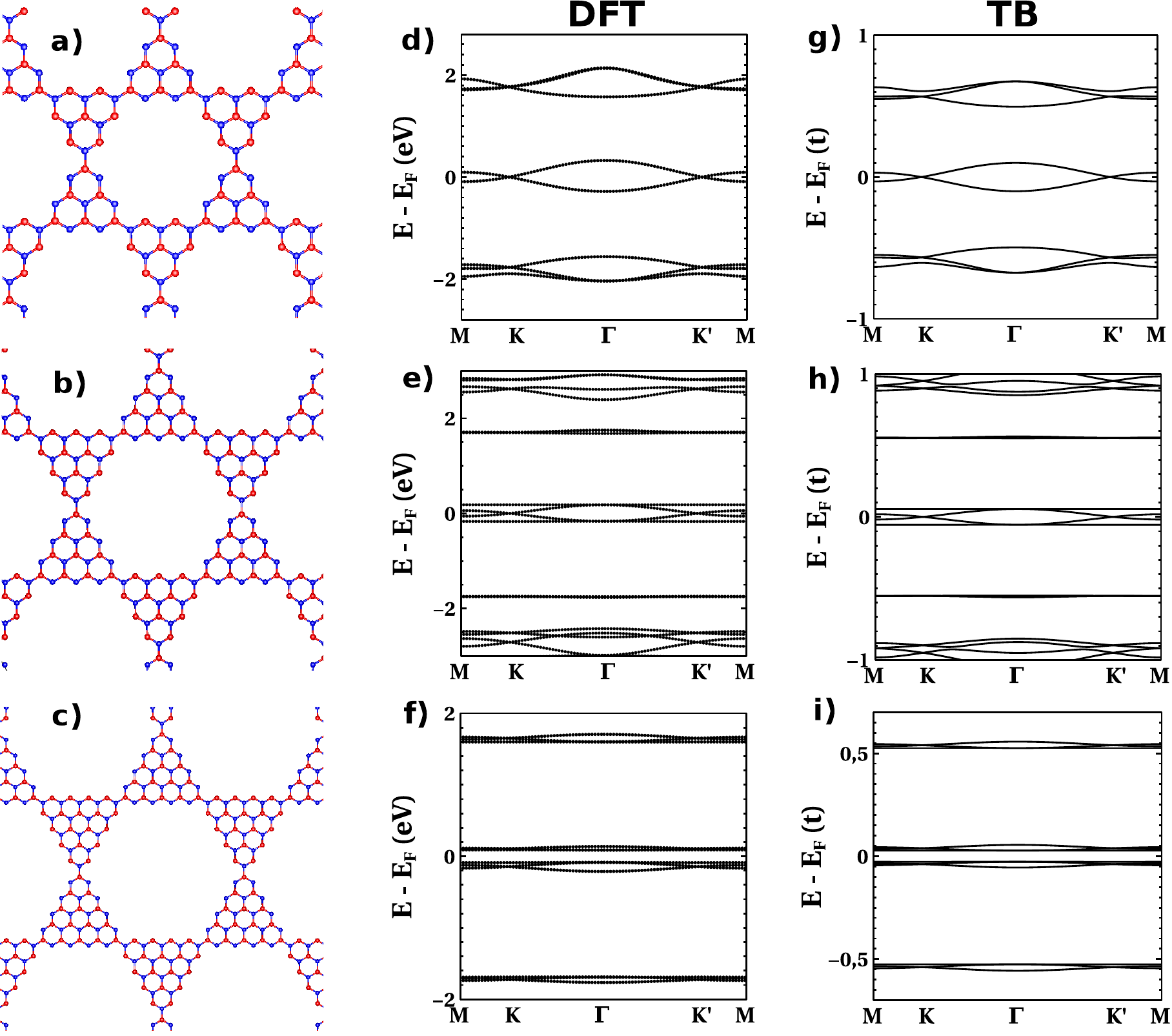}
\caption{
(a-c) Atomic structure of (a) $[2,2]$-, (b) $[3,3]$- and (c) $[4,4]$triangulene crystals.
Red and blue colors denote the two sublattices. 
(d-f) Spin-unpolarized DFT bands of (d) $[2,2]$-, (e) $[3,3]$- and (f) $[4,4]$triangulene crystals.
(g-i) Energy bands obtained with the TB model, including third neighbor hopping $t_3 = 0.1 t$, for (g) $[2,2]$-, (h) $[3,3]$- and (i) $[4,4]$triangulene crystals.
}
\label{fig:3}
\end{figure*}

We now discuss the energy bands of several $[n_a,n_b]$triangulene 2D crystals, calculated with two methods: spin-unpolarized DFT and TB (see results for $n_a=n_b=2,3,4$ in Fig.~\ref{fig:3}).
The spin-unpolarized DFT calculations enforce non-magnetic solutions, so that interactions do not break time-reversal nor sublattice symmetry. 
Our DFT calculations were carried out with Quantum Espresso\cite{giannozzi2009quantum}, using the Perdew-Burke-Ernzerhof functional\cite{perdew1996generalized}.
The edge carbon atoms were passivated with hydrogen.
The kinetic energy cutoff considered was 30 Ry.
The charge density and potential energy cutoffs used were 700 Ry.
We employed $\vec{k}$-grids of $8 \times 8 \times 1$ for the $[3,3]$triangulene crystal and of $10 \times 10 \times 1$ for the $[2,2]$ and $[4,4]$ cases.
No significant deviations from planarity were found.

Below we show that the resulting low-energy bands are associated with the triangulene zero modes and can be accounted for by the TB calculations, as long as third neighbor hopping is included.
We start by addressing the DFT bands obtained for the $[2,2]$-, $[3,3]$- and $[4,4]$triangulene crystals (Fig.~\ref{fig:3}d-f).
The details of these first-principle calculations can be found in Appendix~\ref{appendix:DFT_details}.
Our results are in agreement with previous work by the group of Feng Liu\cite{zhou20,sethi2021} for the $[2,2]$ and $[4,4]$ cases.

The overall picture of the spin-restricted DFT bands for the $[n,n]$triangulene, with $n=2,3,4$, is the following.
A set of $2(n-1)$ weakly dispersive bands is located around the Fermi energy, well separated from higher/lower energy bands.  
Given that the number of zero modes per unit cell is $2(n-1)$, it can be expected that the wave functions of these bands are mostly made of zero modes. 
In the case of $n=2$ (Fig.~\ref{fig:3}d), the bands are isomorphic to the $\pi$ bands of graphene, but with a {\em smaller} bandwidth of 605 meV. 
The $n=3$ bands, shown in Fig.~\ref{fig:3}e, feature two Dirac cones and, in addition, two flat bands at the maximum and minimum of the graphene-like Dirac bands.
We note that these bands are identical to those obtained in the $p_{x,y}$-orbital honeycomb model\cite{wu2007,wu2008}.
The spectrum of the $[4,4]$triangulene crystal (Fig.~\ref{fig:3}f, see also a zoom in Fig.~\ref{fig:5}g) features a small gap of 0.17 eV, with flat valence/conduction bands and a pair of graphene-like bands whose top (bottom) is degenerate with the valence (conduction) flat band at the $\Gamma$ point.
We note that the same set of bands was found in crystalline networks where topological zero-energy states are hosted at the junctions of a network of graphene nanoribbons\cite{tamaki2020}.

The appearance of non-dispersive bands at finite energy is quite peculiar, as non-bonding states in bipartite lattices usually correspond to sublattice-polarized modes with $E=0$.
Here, the sublattice-polarized $E=0$ states seem to hybridize, forming pairs of bonding-antibonding states that lead to electron-hole symmetric bands, some of which being flat bands with finite energy.
The origin of the peculiar dispersion of these low-energy bands, expected to be all flat at $E=0$ in the nearest-neighbor TB model, is addressed in the next section.

The dispersion of the low-energy bands obtained with DFT suggests that the zero modes of adjacent triangulenes do hybridize, so that the nearest neighbor TB approximation fails. 
Second neighbor hopping does not couple states hosted at different sublattices, including the zero modes of adjacent triangulenes, and, in addition, leads to bands that break electron-hole symmetry.
Therefore, it is natural to consider a third neighbor hopping, $t_3$.  
Our calculations with $t_3=0.1 t$ show energy bands in qualitative agreement with those of DFT for the $[2,2]$-, $[3,3]$- and $[4,4]$triangulene crystals (Fig.~\ref{fig:3}g-i).
If we take $t_3=0$ (not shown), the $n_a+n_b-2$ low-energy bands become flat, with $E=0$.
Thus, we conclude that third neighbor hopping accounts for the observed dispersion and is the key energy scale that controls inter-triangulene hybridization.

%%%%%%%%%%%%%%%%%%%%%%%%%%%%%%%%%%%%%%%%%%%%%%%%%%%%%%%%%%%%%%%%%%%%%%%%%%%%%

\section{Effective low-energy theory}

The TB results of the previous section illustrate that $[n_a,n_b]$triangulene 2D crystals provide a platform to generate a wide class of non-trivial energy bands, including graphene-like Dirac dispersion, $p_{x,y}$-orbital honeycomb physics and a gapped system with flat valence and conduction bands.
The emergence of these peculiar results calls for a deeper comprehension.

\begin{figure}
\includegraphics[width=0.9\linewidth]{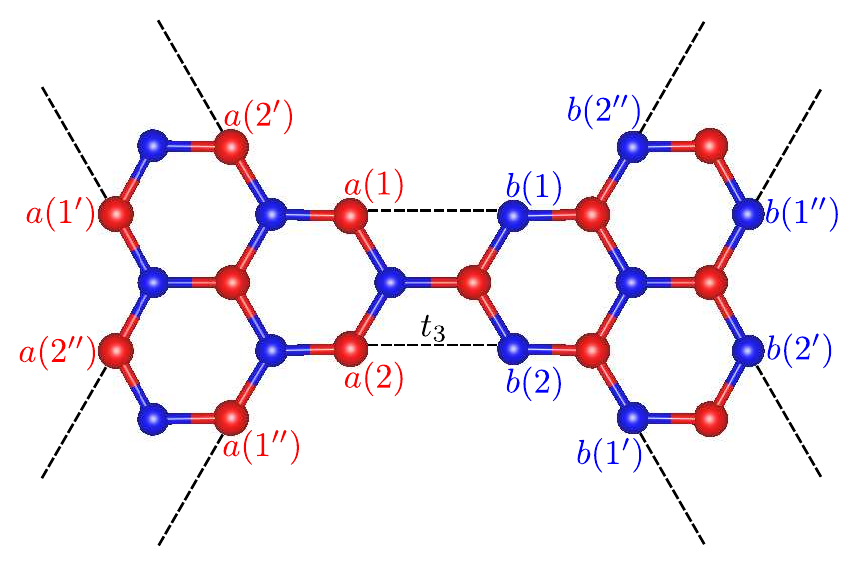}
\caption{
Scheme illustrating how third neighbor hopping $t_3$ leads to hybridization of zero-energy states in adjacent triangulenes via three pairs of atomic dimers.
The zero mode wave functions at atomic sites labeled with the same numerals ($j=1,2$) but with different number of primes are related by $C_3$ symmetry.
}
\label{fig:4}
\end{figure} 

In order to ascertain the ultimate origin of the peculiar properties of the low-energy bands of $[n_a,n_b]$triangulenes, we define an effective Hamiltonian in terms of the $n_a+n_b-2$ zero modes in the unit cell, calculated with the TB model with $t_3=0$\footnote{See Ref.\cite{Catarina2022} for a similar derivation applied to a toy model for 1D crystals of [3]triangulenes.}.
These zero modes are split into two groups consisting of $n_a-1$ $|a\rangle$ and $n_b-1$ $|b\rangle$ states localized in the $A$ and $B$ sublattices of adjacent triangulenes, respectively.
Third neighbor hopping connects $|a\rangle$ and $|b\rangle$ at three pairs of atomic dimers at the corners of the triangulenes (Fig.~\ref{fig:4}).
On the other hand, $t_3$ does not couple zero modes of the same triangulene, as they belong to the same sublattice.
The resulting effective model defines a honeycomb lattice with $n_a+n_b-2$ states per unit cell, whose Bloch Hamiltonian can be written as:
\begin{equation}
\mathcal{H}_\text{eff}(\vec{k}) = 
\begin{pmatrix}
0^{[a]} & \tau (\vec{k}) \\
\tau^\dagger (\vec{k}) & 0^{[b]}
\end{pmatrix},
\label{eq:Heff}
\end{equation}
where $\vec{k}$ denotes the 2D crystal momentum, $0^{[a/b]}$ are square matrices of dimension $n_{a/b}-1$ with all entries equal to zero, and $\tau$ is an $(n_a-1)\times(n_b-1)$ rectangular matrix with entries given by:
\begin{equation}
\tau_{ab}(\vec{k})=\sum_{m=0,1,2} e^{i \vec{k} \cdot \vec{R}_m} t^{(m)}_{ab}.
\label{eq:tau}
\end{equation}
In the previous expression, $\vec{R}_0=(0,0)$ is the null vector associated to the intracell hopping terms, and $\vec{R}_{1,2}$ are the primitive vectors that define the hexagonal lattice of $[n_a,n_b]$triangulenes:
\begin{equation}
\vec{R}_{1,2}=(n_a+n_b+1)\vec{a}_{1,2},
\label{crystalvectors}
\end{equation}
where $\vec{a}_{1,2}$ are the primitive vectors of the graphene hexagonal lattice.
The ($\vec{k}$-independent) hopping matrix elements $t^{(m)}_{ab}$ are defined below.
As usual, the hopping between $A$ and $B$ states of the honeycomb lattice has one intracell (the $m=0$ term of the sum in Eq.~\eqref{eq:tau}) and two intercell (the $m=1,2$ terms of the sum in Eq.~\eqref{eq:tau}) contributions.

The intracell hopping matrix elements can be expressed as:
\begin{equation}
t^{(0)}_{ab}= t_3 \sum_{j=1,2} a(j) b^*(j),
\label{tab}
\end{equation}
where $a(j) = \langle j | a \rangle$ and $b(j) = \langle j | b \rangle$ denote the components of the zero mode wave functions $|a\rangle$ and $|b\rangle$ at the atomic sites $j=1,2$, following the labeling depicted in Fig.~\ref{fig:4}.
Importantly, we can always impose that $a(1) = b(1) \in \mathbb{R}$, from which it follows that $a(2) = b(2) \in \mathbb{R}$.
In the following, we shall assume this gauge fixing.
For that reason, we will use the notation $z$ for cases where we want to label either $a$ or $b$.

\begin{table}
\begin{tabular}{ccccc}
$n$ & zero mode & $\omega_z$ & $z(1)$ & $z(2)$ \\ \hline
2 & $z_0$ & 0 & $\frac{1}{\sqrt{6}}$ & $\frac{-1}{\sqrt{6}}$ \\ \hline
3 & $z_+$ & $+\frac{2\pi}{3}$ & $\frac{1}{\sqrt{11}}$ & $\frac{-1}{\sqrt{11}}$ \\
3 & $z_-$ & $-\frac{2\pi}{3}$ & $\frac{1}{\sqrt{11}}$ & $\frac{-1}{\sqrt{11}}$\\ \hline
4 & $z_0$ & 0 & $\frac{1}{\sqrt{12}}$ & $\frac{-1}{\sqrt{12}}$ \\
4 & $z_+$ & $+\frac{2\pi}{3}$ & $\frac{1}{\sqrt{21}}$ & $\frac{-1}{\sqrt{21}}$\\
4 & $z_-$ & $-\frac{2\pi}{3}$ & $\frac{1}{\sqrt{21}}$ & $\frac{-1}{\sqrt{21}}$ \\ \hline
\end{tabular}
\caption{
From left to right: size of the triangulene, labels for the corresponding $n-1$ zero-energy modes, phase that each zero mode wave function acquires upon the action of the $R_{2\pi/3}$ rotation operator (see Eq.~\eqref{eq:C3basis} and Fig.~\ref{fig:2}b), and values of each zero mode wave function at the binding sites $j=1,2$.
The label $z$ stands for either $a$ or $b$.
}
\label{tab:1}
\end{table}
 
Using the point group symmetry of the triangulenes, the intercell hopping matrix elements can be expressed in terms of the intracell ones.
Given the values of $z(1)$ and $z(2)$, together with the phases $\omega_z$ that each zero mode wave function acquires when applying the $R_{2\pi/3}$ operator (see Table~\ref{tab:1} for $n=2,3,4$), the remaining values of $z(j')$ and $z(j'')$ can determined by making use of the $C_3$ symmetry.
Specifically, we use that $z(j) = e^{i\omega_z} z(j') = e^{-i\omega_z} z(j'')$ to obtain:
\begin{equation}
t^{(1)}_{ab} =  t_3 \sum_{j'=1',2'} a(j') b^*(j') = e^{i(\omega_b-\omega_a)} t^{(0)}_{ab}
\end{equation}
and
\begin{equation}
t^{(2)}_{ab} =  t_3 \sum_{j''=1'',2''} a(j'') b^*(j'') = e^{i(\omega_a-\omega_b)} t^{(0)}_{ab},
\end{equation}
which fully clarifies how to build the effective Hamiltonian of Eq.~\eqref{eq:Heff}.

The effective Hamiltonian described above is the {\em minimal} TB model that is expected to capture the $n_a+n_b-2$ narrow bands of $[n_a,n_b]$triangulenes.
This minimal model describes a honeycomb lattice whose unit cell has $n_a-1$ and $n_b-1$ states in the $A$ and $B$ sublattice-type sites, respectively.
We can think of the different states of the same sublattice as a pseudospin degree of freedom with $C_3$ symmetry, thereby generalizing the honeycomb model of graphene to the case of fermions with an additional internal structure.
The magnitude of the effective hopping is controlled by the weight of the zero modes at the intracell binding sites $j=1,2$.
The intercell hopping matrix elementns depend on the phase differences $\omega_a-\omega_b$.
When nonzero, we can interpret the phases $\omega_z$ as Peierls phases associated to a pseudospin-dependent magnetic field.
Importantly, the model must still be time-reversal invariant.

\begin{figure*}
\includegraphics[width=0.9\linewidth]{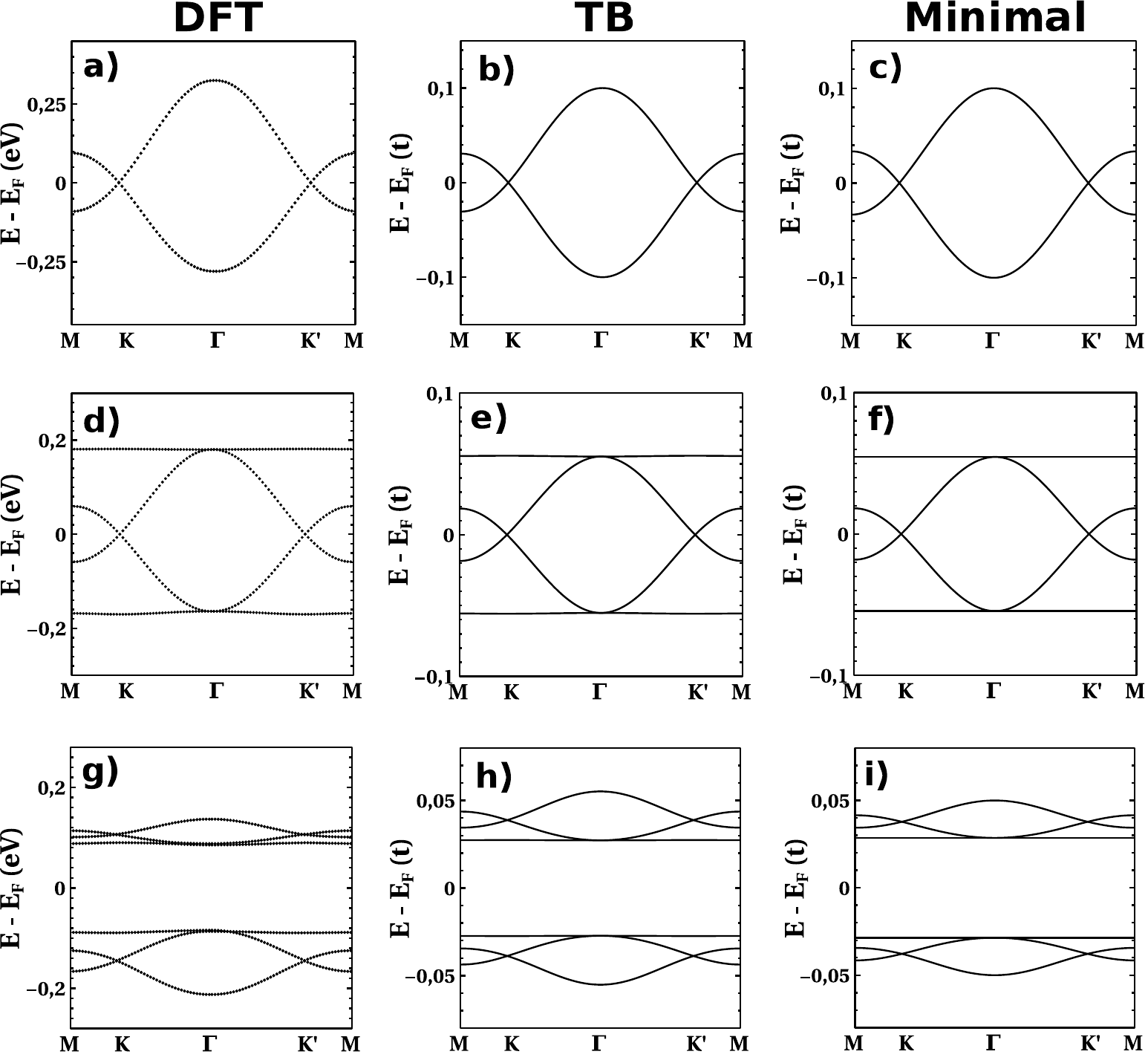}
\caption{
Energy bands of $[n,n]$triangulene crystals, for (a-c) $n=2$, (d-f) $n=3$ and (g-i) $n=4$, obtained with (a,d,g) spin-unpolarized DFT, (b,e,g) the full TB model and (c,f,i) the minimal TB model defined by Eq.~\eqref{eq:Heff}.
The TB results were obtained with $t_3 = 0.1t$.
}
\label{fig:5}
\end{figure*}

In Fig.\ref{fig:5}c,f,i, we show the energy bands obtained with the minimal TB model for $n_a = n_b = 2,3,4$. taking $t_3 = 0.1t$.
We find an excellent agreement with the full TB model (Fig.\ref{fig:5}b,e,h), which in turn agrees with the spin-unpolarized DFT results (Fig.\ref{fig:5}a,d,g).
Below, we take an in-depth look into the effective Hamiltonian of Eq.~\eqref{eq:Heff}, which allow us to have a better understanding of the band dispersion for $n_a=n_b=2,3,4$, as well as for the non-centrosymmetric case $n_a=2,n_b=3$ (not addressed yet).

\subsection{$[2,2]$triangulene crystal}

The minimal Hamiltonian for the $[2,2]$triangulene crystal is a $2 \times 2$ matrix that is isomorphic to the TB model for a monoatomic honeycomb lattice with one orbital per site and nearest neighbor hopping.
The reason for that is the fact that there is only one zero mode per triangulene, with $\omega_z=0$.
Using that $z(1)=-z(2)=\frac{1}{\sqrt{6}}$ (see Table~\ref{tab:1}), we get $t^{(0)}_{ab} = t^{(1)}_{ab} = t^{(2)}_{ab} = t_3/3$.
Thus, we obtain the following effective Bloch Hamiltonian:
\begin{equation}
\mathcal{H}^{[2,2]}_\text{eff}(\phi_1,\phi_2) = \frac{t_3}{3}
\begin{pmatrix}
0 & f (\phi_1,\phi_2) \\
f^* (\phi_1,\phi_2) & 0
\end{pmatrix},
\label{eq:H22}
\end{equation}
with
\begin{equation}
f (\phi_1,\phi_2) = 1 + e^{i\phi_1} + e^{i\phi_2}
\end{equation}
and
\begin{equation}
\phi_{1,2} = \vec{k} \cdot \vec{R}_{1,2}.
\end{equation}
The corresponding energy bands are given by:
\small
\begin{align}
\epsilon^{[2,2]}_\pm (\phi_1,\phi_2) &= \pm \frac{t_3}{3} \left| f (\phi_1,\phi_2) \right| \\
&= \pm \frac{t_3}{3} \sqrt{ 3 + 2 \cos \phi_1 + 2 \cos \phi_2 + 2 \cos(\phi_1 - \phi_2) }.
\end{align}
\normalsize

The Hamiltonian obtained in Eq.~\eqref{eq:H22} is identical to the TB model for the $p_z$ orbitals of graphene, but with an effective hopping given by:
\begin{equation}
 t^{[2,2]}_\textrm{eff} = \frac{t_3}{3},
 \label{teff2}
\end{equation}
which leads to an energy bandwidth of $6t^{[2,2]}_\textrm{eff} = 2t_3$.
The bandwidth obtained by our DFT calculations is around $0.6$ eV, which implies $t_3 \simeq 0.3$ eV, in line with the values from the literature~\cite{Tran2017}.
A Taylor expansion of Eq.~\eqref{eq:H22} around the $K$ and $K'$ points yields Dirac cones with Fermi velocity $\hbar v_F^{[2,2]} = \frac{5}{2} t_3 d$, where $d$ denotes the carbon-carbon distance.

In short, we see that third neighbor hopping in $[2,2]$triangulene crystals produces an intermolecular hybridization of the zero modes that leads to the formation of an {\em artificial graphene} lattice, with a narrower bandwidth governed by $t_3$, instead of by the nearest neighbor hopping.

\subsection{$[2,3]$triangulene crystal}

\begin{figure}
\includegraphics[width=0.9\linewidth]{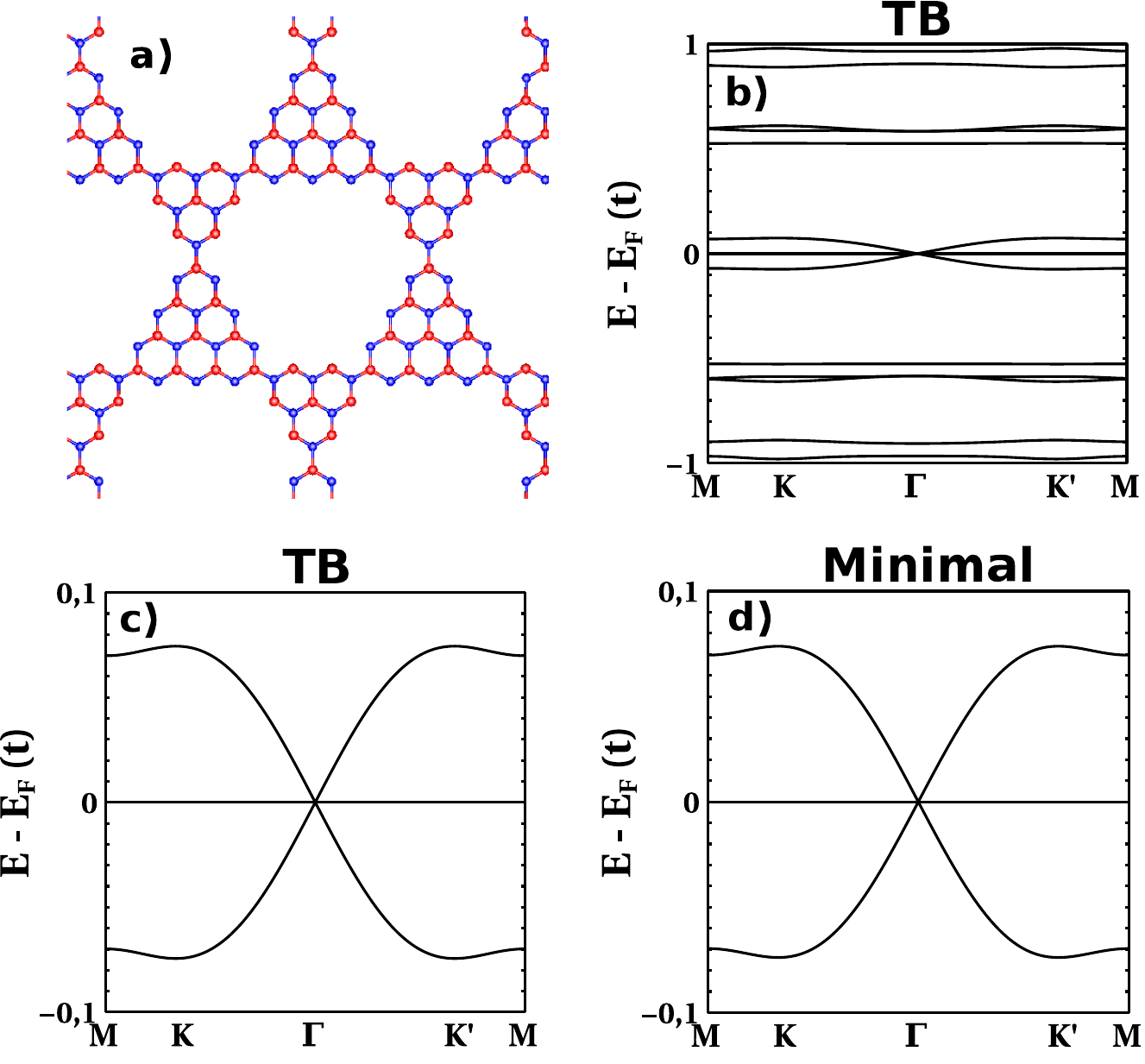}
\caption{
(a) Atomic structure of the $[2,3]$triangulene crystal.
(b) Energy bands of the $[2,3]$triangulene crystal, obtained with the full TB model, using $t_3 = 0.1t$.
(c) Zoom of panel (b) onto the three lowest-energy bands.
(d) Energy bands of the $[2,3]$triangulene crystal, obtained with the minimal TB model, taking $t_3 = 0.1t$.
}
\label{fig:6}
\end{figure}

The $[2,3]$triangulene is the smallest non-centrosymmetric crystal (Fig.~\ref{fig:6}a).
It has a sublattice imbalance of $|N_A-N_B|=1$ that accounts for the presence of one flat band at $E=0$.
Within the minimal model, it has three states per unit cell.
Using the results of Table~\ref{tab:1}, the effective Bloch Hamiltonian can be written as:
\begin{widetext}
\begin{equation}
\mathcal{H}^{[2,3]}_\text{eff} (\phi_1,\phi_2) =
\begin{pmatrix}
0 & \mathcal{F}^{[2,3]}_{a_0,b_+} (\phi_1,\phi_2)  & \mathcal{F}^{[2,3]}_{a_0,b_-} (\phi_1,\phi_2) \\
{\mathcal{F}^{[2,3]}_{a_0,b_+} }^* (\phi_1,\phi_2) & 0 & 0 \\
{\mathcal{F}^{[2,3]}_{a_0,b_-} }^* (\phi_1,\phi_2) & 0 & 0 
\end{pmatrix},
\label{eq:H23}
\end{equation}
\end{widetext}
with
\begin{equation}
\mathcal{F}^{[2,3]}_{a_0,b_\pm} (\phi_1,\phi_2) = \frac{2 t_3}{\sqrt{6 \times 11}} f(\phi_1 \pm \theta, \phi_2 \mp \theta)
\end{equation}
and $\theta = 2\pi/3$.
The corresponding eigenvalues read as:
\begin{equation}
 \epsilon^{[2,3]}_0 (\phi_1,\phi_2) = 0
\end{equation}
and
\small
\begin{align}
 \epsilon^{[2,3]}_\pm (\phi_1,\phi_2) &= \pm \sqrt{ \left| \mathcal{F}^{[2,3]}_{a_0,b_-} (\phi_1,\phi_2) \right|^2 +  \left| \mathcal{F}^{[2,3]}_{a_0,b_-} (\phi_1,\phi_2) \right|^2} \\
 &= \pm \frac{2 t_3}{\sqrt{33}} \sqrt{3 - \cos \phi_1 - \cos \phi_2 - \cos(\phi_1 - \phi_2) }.
\end{align}
\normalsize
In Fig.~\ref{fig:6}d, we plot these energy bands, which are verified to yield an excellent agreement with the low-energy bands of the full TB model (Fig.~\ref{fig:6}b,c).

At the $\Gamma$ point, we have $\phi_1 = \phi_2 = 0$ and therefore $\mathcal{F}^{[2,3]}_{a_0,b_\pm} (0,0) = 0$, which accounts for the triple degeneracy of the bands at that point.
A Taylor expansion of Eq.~\eqref{eq:H23} around $\Gamma$ leads to the following expression for the dispersive bands:
\begin{equation}
\epsilon^{[2,3]}_\pm (\vec{k}) \simeq \pm 6 \sqrt{\frac{3}{11}} t_3 d |\vec{k}|,
\end{equation} 
that accounts for the single Dirac cone.
The corresponding Fermi velocity, $\hbar v_F^{[2,3]} = 6 \sqrt{\frac{3}{11}} t_3 d$, is larger than that obtained in the $[2,2]$triangulene crystal.
It must also be noted that, in the neighborhood of $\Gamma$, the minimal Hamiltonian can be mapped into a spin-1 Dirac model\cite{Mizoguchi2021}.

\subsection{$[3,3]$triangulene crystal}

For $n_a = n_b = 3$, there are two zero modes per triangulene, that we label as $|a_\pm \rangle $ and $|b_\pm \rangle$.
The pairs of zero modes $|z_+ \rangle$ and $|z_- \rangle$ can be distinguished by the following property: $R_{2\pi/3} |z_\pm \rangle = e^{\pm i 2 \pi / 3} |z_\pm \rangle$.
Thus, we can think of the phases $\omega_z = \pm 2\pi/3$ as a pseudospin degree of freedom.

Using the minimal model, the $2 \times 2$ hopping matrix of the effective Bloch Hamiltonian reads as:
\begin{equation}
\tau^{[3,3]} (\phi_1, \phi_2) = 
\begin{pmatrix}
\mathcal{F}^{[3,3]}_{a_+,b_+} (\phi_1,\phi_2) & \mathcal{F}^{[3,3]}_{a_+,b_-} (\phi_1,\phi_2) \\
\mathcal{F}^{[3,3]}_{a_-,b_+} (\phi_1,\phi_2) & \mathcal{F}^{[3,3]}_{a_-,b_-} (\phi_1,\phi_2)
\end{pmatrix},
\end{equation}
where
\begin{equation}
 \mathcal{F}^{[3,3]}_{a_+,b_+} (\phi_1,\phi_2) = \mathcal{F}^{[3,3]}_{a_-,b_-} (\phi_1,\phi_2) = \frac{2t_3}{11} f (\phi_1,\phi_2),
\end{equation}
\begin{equation}
 \mathcal{F}^{[3,3]}_{a_+,b_-} (\phi_1,\phi_2) = \frac{2t_3}{11} f (\phi_1 +\theta,\phi_2 - \theta)
\end{equation}
and
\begin{equation}
 \mathcal{F}^{[3,3]}_{a_-,b_+} (\phi_1,\phi_2) = \frac{2t_3}{11} f (\phi_1 -\theta,\phi_2 + \theta).
\end{equation}
Thus, we see that the {\em pseudospin conserving} terms lead to the same type of matrix elements than in the conventional honeycomb graphene model, whereas the {\em pseudospin flip} terms contain additional $\theta = 2\pi/3$ phases.

Diagonalizing the minimal Hamiltonian, we obtain
\begin{equation}
 \epsilon_{1,\pm}^{[3,3]} (\phi_1,\phi_2) = \pm 3 t_\mathrm{eff}^{[3,3]},
\end{equation}
that accounts for the flat bands at finite energy, and
\begin{equation}
 \epsilon_{2,\pm}^{[3,3]} (\phi_1,\phi_2) = \pm t_\mathrm{eff}^{[3,3]} | f (\phi_1,\phi_2)|,
\end{equation}
that accounts for the graphene-like bands, with an effective hopping $t_\mathrm{eff}^{[3,3]} = 2 t_3 / 11$.
The dispersive graphene-like bands feature Dirac cones at the $K$ and $K'$ points ($\phi_1 = -\phi_2 = \pm 2 \pi /3$), with Fermi velocity $\hbar v_F ^{[3,3]} = \frac{21}{11} t_3 d$, which is smaller than those obtained for the $[2,2]$ and $[2,3]$ cases. 

The emergence of flat bands at finite energy is remarkable. 
These bands are associated to states that are localized around any given hexagonal ring in the honeycomb lattice\cite{wu2007,wu2008,tamaki2020}.  
Every unit cell of the $[3,3]$triangulene crystal must contribute with one state to each of the two flat bands.  
Thus, this makes half a state per triangulene and band. 
Every triangulene participates in three hexagonal rings in a honeycomb lattice. 
Therefore, every triangulene contributes with $1/6$ of state per band to a given hexagonal ring.  
Thus, the six triangulenes that form a ring give rise to one localized state, plus its electron-hole partner.
Using numerical diagonalization for finite size clusters, we have 
verified that, indeed, every hexagonal ring hosts two localized states with energies $\pm 3 t_\mathrm{eff}^{[3,3]}$. 
These states are bonding, in the sense that both sublattices participate, and are thus different from $E=0$ flat bands, which are sublattice-polarized.

\subsection{$[4,4]$triangulene crystal}

In contrast to the previous triangulene 2D crystals, the energy bands of the $[4,4]$triangulene feature a gap at half-filling. 
As in the previous cases, the spin-unpolarized DFT bands close to $E=0$ are well described both by the full TB Hamiltonian and by the minimal low-energy model, as shown in Fig.~\ref{fig:5}g-i.  
The three topmost valence bands, and their conduction band electron-hole partners, are identical to the bands of a Kagome lattice: a flat band that is degenerate at the $\Gamma$ point with the top/bottom of two graphene-like bands. 
The lowest (highest) energy conduction (valence) band is thus dispersionless, which is of particular interest for the emergent field of flat band physics.

Given that, for the $[4,4]$triangulene crystal, the number of zero modes per triangulene (three) is a multiple of the coordination number of the honeycomb lattice (three), it is natural to associate its band gap to the formation of an orbital valence bond solid. 
This idea is further reinforced by the fact that a gapped band structure is  also obtained for the $[7,7]$triangulene crystal, which has six zero modes per triangulene.

\subsection{Overall picture of the single-particle states}

The low-energy bands of $[n_a,n_b]$triangulene crystals are remarkable for several reasons. 
First, they only disperse because of a third neighbor hopping, that would normally play a residual role, yet it is dominant here.  
Second, they feature flat bands at finite energy, which are interesting on their own right.
Third, they provide a generalization of the honeycomb TB model, with a single orbital per site, to a more general class of honeycomb models where fermions have an internal pseudospin degree of freedom.

At this point, we can draw an analogy between this pseudospin degree of freedom and the orbital angular momentum of atoms, $\ell$.  
For atoms, the orbital degeneracies are given by $2\ell+1$, with $\ell=0,1,2,...$, and their symmetry properties are governed by the spherical harmonics. 
For $[n]$triangulenes, the orbital degeneracies are given by $n-1$, with $n=2,3,4,...$, and their symmetry properties are determined by the discrete $R_{2\pi/3}$ operator. 
Very much like the angular momentum wave function of the valence electrons in atoms shapes their electronic properties, the pseudospin of the fermions in $[n_a,n_b]$triangulene crystals dictates the resulting band structures.
In particular, the $[2,2]$ case leads to artificial graphene, $[2,3]$ to the $S=1$ Dirac Hamiltonian, and $[3,3]$ to the $p_{x,y}$-orbital honeycomb model, all of which featuring Dirac cones at the Fermi energy.
The $[4,4]$triangulene, in contrast, is a band insulator, with flat valence and conduction bands.
It must be noted that, even though we have been always considering triangulene 2D crystals at half-filling, other options could be possible if carriers are injected into them, for instance via the application of a gate voltage or through chemical doping\cite{Wang2022}.

Finally, we note that even though the dispersive low-energy bands are narrow, interactions will turn these systems into Mott insulators at half-filling, as we discuss in the next section.

%%%%%%%%%%%%%%%%%%%%%%%%%%%%%%%%%%%%%%%%%%%%%%%%%%%%%%%%%%%%%%%%%%%%%%%%%%%%%

\section{Effect of interactions}

\subsection{General considerations}

We now discuss {\em qualitatively} the effect of interactions. 
Unless otherwise stated, we focus on the half-filling case. 
At the single triangulene level, interactions have a huge impact on the electrons at the zero modes, and they determine the strong intramolecular ferromagnetic exchange that leads to $2S=n-1$, for $[n]$triangulenes.
In the case of triangulene crystals,  Coulomb repulsion competes with inter-triangulene hybridization.
We quantify this competition in terms of the ratio of two energy scales, that we can define for each zero mode. 
On the one hand we have the effective Coulomb interaction, 
\begin{equation}
\tilde U_a=U \sum_i |a(i )|^4.
\end{equation}
On the other hand, the average inter-triangulene hopping energy,
\begin{equation}
K_a=\frac{1}{n_b}\sum_{b} \tau_{ab}(0),
\end{equation}
where $\tau_{ab}(0)$ is the expression given in Eq.~\eqref{eq:tau}, evaluated at the $\Gamma$ point. 
In order to obtain a single figure of merit for a given unit cell, we can then average the ratio $\tilde U_a/K_a$ over the zero modes. 
The results depend both on the values of atomic energy scales $U$ and $t_3$  and on properties of the molecular orbitals, such as their inverse participation ration (IPR), given by $\sum_i |a(i )|^4$, and the hopping matrix elements. 
We find that, for all triangulene systems considered here, this ratio
is systematically much larger than $1$ if we take typical values\cite{Mishra2020,Mishra2021} of $U=1.5 t$ and $t_3=0.1 t$. 
Thus, we expect that, at half filling, triangulene 2D crystals will be Mott insulators. 
Perhaps the only exception could be the $[4,4]$triangulene that has a gap before interactions are included.

We illustrate our point with the case of the $[2,2]$triangulene dimer, for which the IPR of the zero mode is $1/6$, so that the ratio $r\equiv \frac{\tilde{U}}{t^{[2,2]}_{\rm eff}}=  \frac{U/6}{t_3/3}=\frac{U}{2t_3}$. For $U=1.5 t$ and $t_3=0.1 t$, we have $r\simeq 7.5$, clearly in the insulating side of the metal-insulator transition predicted for $r>4.5$ for the Hubbard model at half filling in the honeycomb lattice\cite{sorella92}.  In the Mott insulating regime, with the charge fluctuations frozen, we expect the $[2,2]$triangulene crystal to behave like a $S=1/2$ honeycomb lattice with antiferromagnetic exchange
\begin{equation}
J^{[2]} =4\frac{(t^{[2,2]}_{\rm eff})^2}{\tilde{U}}=6\times \frac{4}{9} \frac{t_3^2}{U} \simeq 48 {\rm meV},
\end{equation}
for $t=2.7 eV$, $U=1.5 t$ and $t_3=0.1 t$. 
As we shall discuss in a forthcoming work\cite{David22}, in addition to the kinetic exchange, there are other contributions to the antiferromagnetic exchange in this system.

\subsection{Spin-polarized DFT calculations for the $[3,3]$triangulene crystal}

\begin{figure}[t]
\includegraphics[width=0.9\linewidth]{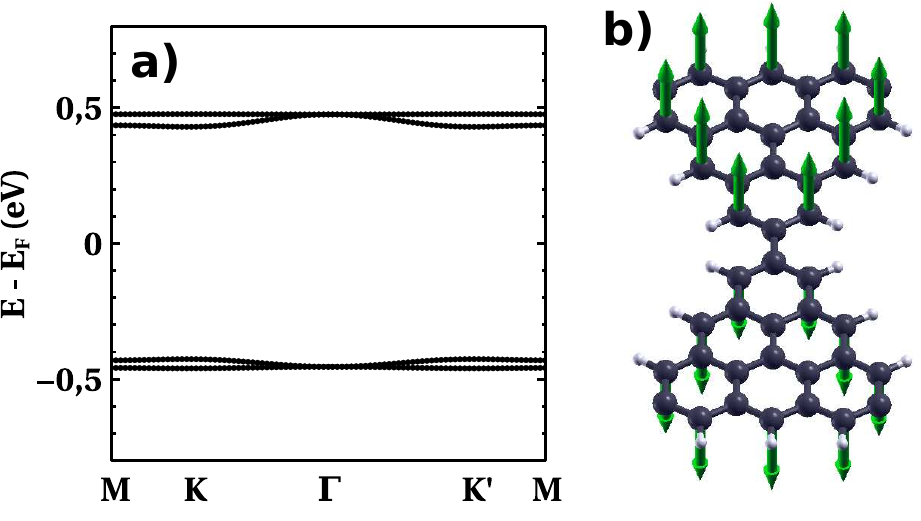}
\caption{
a) Energy bands, computed with spin-polarized DFT, for the $[3,3]$triangulene crystal.
b) Computed magnetic moment density, showing antiferromagnetic order.
}
\label{fig:8}
\end{figure}

We now discuss the results of our spin-polarized DFT calculations on the $[3,3]$triangulene crystal (see details in Appendix~\ref{appendix:DFT_details}). 
They provide a realistic assessment of the emergence of local moments and their exchange interactions but can only describe broken symmetry 
magnetic states, and therefore cannot account for quantum disordered spin phases that are known to occur for 1D chains of triangulene crystals\cite{Mishra2021}.   
The results are shown in Fig.~\ref{fig:8} for the lowest energy magnetic configuration, where the triangulenes have local moments with opposite magnetizations.  
The four lowest energy bands clearly show that a very large gap has opened where the Dirac bands once stood.  
Additionally, from the magnetic moments distribution across the molecules, shown in Fig.~\ref{fig:8}b, it is apparent that local moments are hosted by the zero modes.
We note that spin-polarized DFT results for the $[2,2]$triangulene crystal obtained by Zhou and Liu\cite{zhou20}, also feature an antiferromagnetic insulator. 
Therefore, the available spin-polarized DFT results support the picture that $[n,n]$triangulenes are Mott insulators with antiferromagnetically coupled local moments.

From our calculations we can infer the value of the intermolecular exchange $J^{[3]}$. 
For that matter, we compare the energy difference between the ferromagnetic and the antiferromagnetic solutions obtained with DFT, with the result of a classical Heisenberg model. 
The energy difference per unit cell is given by $6 J^{[3]} S(S+1)=E_{FM}-E_{AFM}=159.2$meV.  
Using $S=1$ we infer $J^{[3]}=26.5$meV. 
This analysis, that overlooks additional contributions such as biquadratic exchange, is to be compared with the result of $J=14$meV, recently inferred for the $S=1$ triangulene dimer, in the same approximation\cite{Mishra2020}.  
 
\subsection{Spin phases in honeycomb crystals}

We now briefly discuss the type of interacting ground states that can be expected to arise for the  triangulene crystals at half filling. 
In the case of $[2,2]$-crystals,   the relevant reference are the  quantum Monte Carlo calculations for the Hubbard model in the half-limit case. These show a transition to a Neel ordered Mott insulating phase for $U>4.2 t$.  The existence, or else, of a spin liquid phase has been studied extensively using Heisenberg spin models\cite{gong2013}. In the case of $[2,3]$ at half filling, Lieb theorem\cite{Lieb89} ensures a ground state  with $S=1/2$ per unit cell. According to Lieb theorem the 
non-centrosymmetryc  crystals $[n_a,n_b]$ will have a spin $2S=|n_a-n_b|$ per unit cell, and  
 will be either ferromagnetic or ferrimagnetic.

The $S=1$ Heisenberg model in the honeycomb lattice has also been studied using density matrix renormalization group (DMRG) calculations in finite size clusters\cite{gong2015}. 
These calculations predict  a Neel phase with long-range order, unless  second-neighbor antiferromagnetic interactions, that promote frustration,  are significant.   For bipartite lattices, second-neighbor exchange coupling would be ferromagnetic. Therefore,  a Neel  phase is  likely to occur for the $[3,3]$triangulene crystal at half filling.

In the case of the $[4,4]$ triangulene crystal, the relevant spin model is the $S=3/2$ Heisenberg Hamiltonian in the honeycomb lattice with antiferromagnetic interactions.  Interestingly, this Hamiltonian is not far from the AKLT model for $S=3/2$, that  contains Heiseberg (bilinear) terms as well as  biquadratic and bicubic spin couplings\cite{Affleck1987}. Importantly, the AKLT model can be solved exactly and features a spin valence bond-solid (VBS).  The ground state of the AKLT model 
is a universal resource for measurement based quantum computing\cite{Wei2011} (MBQC). Importantly, the AKLT model has
 a gap in the excitation spectrum\cite{Garcia-Saez2013,pomata20,Lemm20}, so that cooling down a system described with the AKLT model at  low enough temperatures would be enough to prepare a universal resource for MBQC.
 
 The  AKLT point is a singular point in a class of models, with bilinear and biquadratic couplings (BLBQ) and even bicubic\cite{Affleck1987} models in the honeycomb case.
 For 1D,  VBS phase is  robust in a region of parameters containing both  AKLT and  the pure Heisenberg limit.  
 For the 2D honeycomb AKLT model, finite-size diagonalisation of the spin model\cite{ganesh2011}
 found a VBS to Neel transition when AKLT is linearly transformed into a  Heisenberg model, but the precise location of the transition is hard to establish due to finite-size effects. In any event, the interesting question of whether the $[4,4]$triangulene crystal provides a realization of the VBS in two dimensions remains open.  We note that the VBS phase  predicted for the $S=1$ AKLT in 1D, and it has been observed for $[3]$triangulene chains\cite{Mishra2021}.  

Deviations from half-filling, introducing carriers in the Mott insulating phases, will very likely bring very interesting electronic phases, as it happens in the case of cuprates and in the case of twisted bilayer graphene\cite{cao2018b}.  This matter deserves further study. 

\section{Summary and conclusions}
We  have shown that graphene triangulenes  are ideal building blocks for bottom-up design of carbon-based 2D  honeycomb crystals with both narrow and flat bands at the Fermi energy. 
Our study permits to set the rules for the rational design of the bands, including the number of narrow bands, and the presence of flat bands either at the Fermi energy or nearby.   The design principles for honeycomb crystals with a  unit cell made of two triangulenes, of dimensions $n_a,n_b$ (see Figure \ref{fig:1}) are:
\begin{enumerate}
\item Isolated $[n]$triangulenes host $n-1$ zero modes that are localized in one sublattice (either $a$ or $b$, depending on the orientation) (see Figure \ref{fig:1})
\item The zero modes vanish identically at the inter-triangulene binding sites. As a result intermolecular hybridization is governed by third neighbor hopping $t_3$.
\item  Centrosymmetryc $[n,n]$triangulene crystals have $2(n-1)$ narrow bands. 
For $n\geq3$ they always feature pairs of electron-hole conjugate low energy flat bands, corresponding to states localized in supramolecular hexagonal rings. These flat bands are very different from $E=0$ bands, as they are not sub-lattice polarized and they feature intermolecular hybridization.
\item Non centrosymmetric $[n_a,n_b]$triangulene crystals, with $n_a\neq n_b$,   feature  $|n_a-n_b|$ flat bands at $E=0$, because of the bipartite character of the lattice\cite{Sutherland1986,Ortiz19}
\end{enumerate}

The results above are validated comparing spin-unpolarized DFT calculations with tight-binding Hamiltonians at two levels: a full-lattice model that includes first and third neighbor hopping and a minimal  tight-binding model where only the zero modes of the triangulenes are included.  Using the latter approach we are able to derive analytical expressions for the energy bands of the $[2,2]$,$[2,3]$, and $[3,3]$  triangulene crystals.

We have also addressed the effect of electron-electron interactions on these crystals. Both analytical estimates of the relevant energy scales and spin-polarized DFT calculations permit to anticipate that interactions will dominate the electronic properties of these crystals. At half-filling these crystals are expected to be Mott insulators, with their low energy physics governed by the spin degrees of freedom.  Application of Lieb theorem\cite{Lieb89} valid for the Hubbard model in bipartite lattices at half filling permits to anticipate that compensated ($[n,n]$) triangulene crystals have a ground state with $S=0$, and therefore antiferromagnetic intermolecular exchange.  We have estimated the magnitude of this super-exchange for the $[2,2]$ and $[3,3]$ crystals, and found very large values, in the range of tens of meV. For non-centrosymmetric $[n_a,n_b]$ crystals with $n_a\neq n_b$ Lieb theorem predicts that the ground state will have $2S=n_a-n_b$ per unit cell. Thus, these crystals will be ferromagnets or ferrimagnets.

Doping these Mott insulators away from half-filling seems   an almost certain recipe to discover non-trivial correlated electronic phases.  We hope our work, together with recent work showing the great interest of triangulene one dimensional structures\cite{Mishra2021} will motivate experimental work to figure out synthetic routes that permit to create triangulene 2D crystals, and the exploration of their electronic properties in the lab.

{\em Acknowledgements}
We acknowledge financial support 
from the Ministry of Science and Innovation of Spain (grant No. PID2019-109539GB-41), 
from Generalitat Valenciana (grant No. Prometeo2021/017) and from Fundacao Para a Ciencia e
a Tecnologia, Portugal (grant No. PTDC/FIS-MAC/2045/2021).
R. Ortiz  acknowledges funding fromGeneralitat
Valenciana and Fondo Social Europeo (Grant No.
ACIF/2018/175).
G. Catarina acknowledges financial support from Funda\c{c}\~{a}o para a Ci\^{e}ncia e a Tecnologia (Grant No. SFRH/BD/138806/2018).

%%%%%%%%%%%%%%%%%%%%%%%%%%%%%%%%%%%%%%%%%%%%%%%%%%%%%%%%%%%%%%%%%%%%%%%%%%%%%
%Appendix

%\appendix

%\section{Details of DFT calculations}
%\label{appendix:DFT_details}

%%%%%%%%%%%%%%%%%%%%%%%%%%%%%%%%%%%%%%%%%%%%%%%%%%%%%%%%%%%%%%%%%%%%%%%%%%%%%
%Bibliography
%\bibliographystyle{apsrev4-2}
%\bibliographystyle{apsrev4}
\bibliography{triang2D}{}

\begin{thebibliography}{56}
\expandafter\ifx\csname natexlab\endcsname\relax\def\natexlab#1{#1}\fi
\expandafter\ifx\csname bibnamefont\endcsname\relax
  \def\bibnamefont#1{#1}\fi
\expandafter\ifx\csname bibfnamefont\endcsname\relax
  \def\bibfnamefont#1{#1}\fi
\expandafter\ifx\csname citenamefont\endcsname\relax
  \def\citenamefont#1{#1}\fi
\expandafter\ifx\csname url\endcsname\relax
  \def\url#1{\texttt{#1}}\fi
\expandafter\ifx\csname urlprefix\endcsname\relax\def\urlprefix{URL }\fi
\providecommand{\bibinfo}[2]{#2}
\providecommand{\eprint}[2][]{\url{#2}}

\bibitem[{\citenamefont{Hart et~al.}(2015)\citenamefont{Hart, Duarte, Yang,
  Liu, Paiva, Khatami, Scalettar, Trivedi, Huse, and Hulet}}]{Hart2015}
\bibinfo{author}{\bibfnamefont{R.~A.} \bibnamefont{Hart}},
  \bibinfo{author}{\bibfnamefont{P.~M.} \bibnamefont{Duarte}},
  \bibinfo{author}{\bibfnamefont{T.-L.} \bibnamefont{Yang}},
  \bibinfo{author}{\bibfnamefont{X.}~\bibnamefont{Liu}},
  \bibinfo{author}{\bibfnamefont{T.}~\bibnamefont{Paiva}},
  \bibinfo{author}{\bibfnamefont{E.}~\bibnamefont{Khatami}},
  \bibinfo{author}{\bibfnamefont{R.~T.} \bibnamefont{Scalettar}},
  \bibinfo{author}{\bibfnamefont{N.}~\bibnamefont{Trivedi}},
  \bibinfo{author}{\bibfnamefont{D.~A.} \bibnamefont{Huse}}, \bibnamefont{and}
  \bibinfo{author}{\bibfnamefont{R.~G.} \bibnamefont{Hulet}},
  \bibinfo{journal}{Nature} \textbf{\bibinfo{volume}{519}},
  \bibinfo{pages}{211} (\bibinfo{year}{2015}), ISSN \bibinfo{issn}{1476-4687}.

\bibitem[{\citenamefont{Mazurenko et~al.}(2017)\citenamefont{Mazurenko, Chiu,
  Ji, Parsons, Kan\'{a}sz-Nagy, Schmidt, Grusdt, Demler, Greif, and
  Greiner}}]{Mazurenko2017}
\bibinfo{author}{\bibfnamefont{A.}~\bibnamefont{Mazurenko}},
  \bibinfo{author}{\bibfnamefont{C.~S.} \bibnamefont{Chiu}},
  \bibinfo{author}{\bibfnamefont{G.}~\bibnamefont{Ji}},
  \bibinfo{author}{\bibfnamefont{M.~F.} \bibnamefont{Parsons}},
  \bibinfo{author}{\bibfnamefont{M.}~\bibnamefont{Kan\'{a}sz-Nagy}},
  \bibinfo{author}{\bibfnamefont{R.}~\bibnamefont{Schmidt}},
  \bibinfo{author}{\bibfnamefont{F.}~\bibnamefont{Grusdt}},
  \bibinfo{author}{\bibfnamefont{E.}~\bibnamefont{Demler}},
  \bibinfo{author}{\bibfnamefont{D.}~\bibnamefont{Greif}}, \bibnamefont{and}
  \bibinfo{author}{\bibfnamefont{M.}~\bibnamefont{Greiner}},
  \bibinfo{journal}{Nature} \textbf{\bibinfo{volume}{545}},
  \bibinfo{pages}{462} (\bibinfo{year}{2017}), ISSN \bibinfo{issn}{1476-4687}.

\bibitem[{\citenamefont{Blatt and Roos}(2012)}]{blatt2012}
\bibinfo{author}{\bibfnamefont{R.}~\bibnamefont{Blatt}} \bibnamefont{and}
  \bibinfo{author}{\bibfnamefont{C.~F.} \bibnamefont{Roos}},
  \bibinfo{journal}{Nature Physics} \textbf{\bibinfo{volume}{8}},
  \bibinfo{pages}{277} (\bibinfo{year}{2012}).

\bibitem[{\citenamefont{Hensgens et~al.}(2017)\citenamefont{Hensgens, Fujita,
  Janssen, Li, Van~Diepen, Reichl, Wegscheider, Das~Sarma, and
  Vandersypen}}]{Hensgens2017}
\bibinfo{author}{\bibfnamefont{T.}~\bibnamefont{Hensgens}},
  \bibinfo{author}{\bibfnamefont{T.}~\bibnamefont{Fujita}},
  \bibinfo{author}{\bibfnamefont{L.}~\bibnamefont{Janssen}},
  \bibinfo{author}{\bibfnamefont{X.}~\bibnamefont{Li}},
  \bibinfo{author}{\bibfnamefont{C.~J.} \bibnamefont{Van~Diepen}},
  \bibinfo{author}{\bibfnamefont{C.}~\bibnamefont{Reichl}},
  \bibinfo{author}{\bibfnamefont{W.}~\bibnamefont{Wegscheider}},
  \bibinfo{author}{\bibfnamefont{S.}~\bibnamefont{Das~Sarma}},
  \bibnamefont{and} \bibinfo{author}{\bibfnamefont{L.~M.~K.}
  \bibnamefont{Vandersypen}}, \bibinfo{journal}{Nature}
  \textbf{\bibinfo{volume}{548}}, \bibinfo{pages}{70} (\bibinfo{year}{2017}),
  ISSN \bibinfo{issn}{1476-4687}.

\bibitem[{\citenamefont{Mortemousque et~al.}(2021)\citenamefont{Mortemousque,
  Chanrion, Jadot, Flentje, Ludwig, Wieck, Urdampilleta, B\"{a}uerle, and
  Meunier}}]{Mortemousque2021}
\bibinfo{author}{\bibfnamefont{P.-A.} \bibnamefont{Mortemousque}},
  \bibinfo{author}{\bibfnamefont{E.}~\bibnamefont{Chanrion}},
  \bibinfo{author}{\bibfnamefont{B.}~\bibnamefont{Jadot}},
  \bibinfo{author}{\bibfnamefont{H.}~\bibnamefont{Flentje}},
  \bibinfo{author}{\bibfnamefont{A.}~\bibnamefont{Ludwig}},
  \bibinfo{author}{\bibfnamefont{A.~D.} \bibnamefont{Wieck}},
  \bibinfo{author}{\bibfnamefont{M.}~\bibnamefont{Urdampilleta}},
  \bibinfo{author}{\bibfnamefont{C.}~\bibnamefont{B\"{a}uerle}},
  \bibnamefont{and} \bibinfo{author}{\bibfnamefont{T.}~\bibnamefont{Meunier}},
  \bibinfo{journal}{Nature Nanotechnology} \textbf{\bibinfo{volume}{16}},
  \bibinfo{pages}{296} (\bibinfo{year}{2021}), ISSN \bibinfo{issn}{1748-3395}.

\bibitem[{\citenamefont{Salfi et~al.}(2016)\citenamefont{Salfi, Mol, Rahman,
  Klimeck, Simmons, Hollenberg, and Rogge}}]{Salfi2016}
\bibinfo{author}{\bibfnamefont{J.}~\bibnamefont{Salfi}},
  \bibinfo{author}{\bibfnamefont{J.~A.} \bibnamefont{Mol}},
  \bibinfo{author}{\bibfnamefont{R.}~\bibnamefont{Rahman}},
  \bibinfo{author}{\bibfnamefont{G.}~\bibnamefont{Klimeck}},
  \bibinfo{author}{\bibfnamefont{M.~Y.} \bibnamefont{Simmons}},
  \bibinfo{author}{\bibfnamefont{L.~C.~L.} \bibnamefont{Hollenberg}},
  \bibnamefont{and} \bibinfo{author}{\bibfnamefont{S.}~\bibnamefont{Rogge}},
  \bibinfo{journal}{Nature Communications} \textbf{\bibinfo{volume}{7}},
  \bibinfo{pages}{11342} (\bibinfo{year}{2016}), ISSN
  \bibinfo{issn}{2041-1723}.

\bibitem[{\citenamefont{Garc\'{i}a-Mart\'{i}nez and
  Fern\'{a}ndez-Rossier}(2019)}]{GarciaMartinez2019}
\bibinfo{author}{\bibfnamefont{N.~A.} \bibnamefont{Garc\'{i}a-Mart\'{i}nez}}
  \bibnamefont{and}
  \bibinfo{author}{\bibfnamefont{J.}~\bibnamefont{Fern\'{a}ndez-Rossier}},
  \bibinfo{journal}{Phys. Rev. Research} \textbf{\bibinfo{volume}{1}},
  \bibinfo{pages}{033173} (\bibinfo{year}{2019}).

\bibitem[{\citenamefont{Kennes et~al.}(2021)\citenamefont{Kennes, Claassen,
  Xian, Georges, Millis, Hone, Dean, Basov, Pasupathy, and Rubio}}]{kennes2021}
\bibinfo{author}{\bibfnamefont{D.~M.} \bibnamefont{Kennes}},
  \bibinfo{author}{\bibfnamefont{M.}~\bibnamefont{Claassen}},
  \bibinfo{author}{\bibfnamefont{L.}~\bibnamefont{Xian}},
  \bibinfo{author}{\bibfnamefont{A.}~\bibnamefont{Georges}},
  \bibinfo{author}{\bibfnamefont{A.~J.} \bibnamefont{Millis}},
  \bibinfo{author}{\bibfnamefont{J.}~\bibnamefont{Hone}},
  \bibinfo{author}{\bibfnamefont{C.~R.} \bibnamefont{Dean}},
  \bibinfo{author}{\bibfnamefont{D.}~\bibnamefont{Basov}},
  \bibinfo{author}{\bibfnamefont{A.~N.} \bibnamefont{Pasupathy}},
  \bibnamefont{and} \bibinfo{author}{\bibfnamefont{A.}~\bibnamefont{Rubio}},
  \bibinfo{journal}{Nature Physics} \textbf{\bibinfo{volume}{17}},
  \bibinfo{pages}{155} (\bibinfo{year}{2021}).

\bibitem[{\citenamefont{Yang et~al.}(2017)\citenamefont{Yang, Bae, Paul,
  Natterer, Willke, Lado, Ferr\'{o}n, Choi, Fern\'{a}ndez-Rossier, Heinrich
  et~al.}}]{Yang2017}
\bibinfo{author}{\bibfnamefont{K.}~\bibnamefont{Yang}},
  \bibinfo{author}{\bibfnamefont{Y.}~\bibnamefont{Bae}},
  \bibinfo{author}{\bibfnamefont{W.}~\bibnamefont{Paul}},
  \bibinfo{author}{\bibfnamefont{F.~D.} \bibnamefont{Natterer}},
  \bibinfo{author}{\bibfnamefont{P.}~\bibnamefont{Willke}},
  \bibinfo{author}{\bibfnamefont{J.~L.} \bibnamefont{Lado}},
  \bibinfo{author}{\bibfnamefont{A.}~\bibnamefont{Ferr\'{o}n}},
  \bibinfo{author}{\bibfnamefont{T.}~\bibnamefont{Choi}},
  \bibinfo{author}{\bibfnamefont{J.}~\bibnamefont{Fern\'{a}ndez-Rossier}},
  \bibinfo{author}{\bibfnamefont{A.~J.} \bibnamefont{Heinrich}},
  \bibnamefont{et~al.}, \bibinfo{journal}{Phys. Rev. Lett.}
  \textbf{\bibinfo{volume}{119}}, \bibinfo{pages}{227206}
  (\bibinfo{year}{2017}).

\bibitem[{\citenamefont{Khajetoorians et~al.}(2019)\citenamefont{Khajetoorians,
  Wegner, Otte, and Swart}}]{khajetoorians2019}
\bibinfo{author}{\bibfnamefont{A.~A.} \bibnamefont{Khajetoorians}},
  \bibinfo{author}{\bibfnamefont{D.}~\bibnamefont{Wegner}},
  \bibinfo{author}{\bibfnamefont{A.~F.} \bibnamefont{Otte}}, \bibnamefont{and}
  \bibinfo{author}{\bibfnamefont{I.}~\bibnamefont{Swart}},
  \bibinfo{journal}{Nature Reviews Physics} \textbf{\bibinfo{volume}{1}},
  \bibinfo{pages}{703} (\bibinfo{year}{2019}).

\bibitem[{\citenamefont{Yang et~al.}(2021)\citenamefont{Yang, Phark, Bae, Esat,
  Willke, Ardavan, Heinrich, and Lutz}}]{Yang2021}
\bibinfo{author}{\bibfnamefont{K.}~\bibnamefont{Yang}},
  \bibinfo{author}{\bibfnamefont{S.-H.} \bibnamefont{Phark}},
  \bibinfo{author}{\bibfnamefont{Y.}~\bibnamefont{Bae}},
  \bibinfo{author}{\bibfnamefont{T.}~\bibnamefont{Esat}},
  \bibinfo{author}{\bibfnamefont{P.}~\bibnamefont{Willke}},
  \bibinfo{author}{\bibfnamefont{A.}~\bibnamefont{Ardavan}},
  \bibinfo{author}{\bibfnamefont{A.~J.} \bibnamefont{Heinrich}},
  \bibnamefont{and} \bibinfo{author}{\bibfnamefont{C.~P.} \bibnamefont{Lutz}},
  \bibinfo{journal}{Nature Communications} \textbf{\bibinfo{volume}{12}},
  \bibinfo{pages}{993} (\bibinfo{year}{2021}), ISSN \bibinfo{issn}{2041-1723}.

\bibitem[{\citenamefont{Cai et~al.}(2010)\citenamefont{Cai, Ruffieux, Jaafar,
  Bieri, Braun, Blankenburg, Muoth, Seitsonen, Saleh, Feng et~al.}}]{cai2010}
\bibinfo{author}{\bibfnamefont{J.}~\bibnamefont{Cai}},
  \bibinfo{author}{\bibfnamefont{P.}~\bibnamefont{Ruffieux}},
  \bibinfo{author}{\bibfnamefont{R.}~\bibnamefont{Jaafar}},
  \bibinfo{author}{\bibfnamefont{M.}~\bibnamefont{Bieri}},
  \bibinfo{author}{\bibfnamefont{T.}~\bibnamefont{Braun}},
  \bibinfo{author}{\bibfnamefont{S.}~\bibnamefont{Blankenburg}},
  \bibinfo{author}{\bibfnamefont{M.}~\bibnamefont{Muoth}},
  \bibinfo{author}{\bibfnamefont{A.~P.} \bibnamefont{Seitsonen}},
  \bibinfo{author}{\bibfnamefont{M.}~\bibnamefont{Saleh}},
  \bibinfo{author}{\bibfnamefont{X.}~\bibnamefont{Feng}}, \bibnamefont{et~al.},
  \bibinfo{journal}{Nature} \textbf{\bibinfo{volume}{466}},
  \bibinfo{pages}{470} (\bibinfo{year}{2010}).

\bibitem[{\citenamefont{Ruffieux et~al.}(2016)\citenamefont{Ruffieux, Wang,
  Yang, S{\'a}nchez-S{\'a}nchez, Liu, Dienel, Talirz, Shinde, Pignedoli,
  Passerone et~al.}}]{Ruffieux16}
\bibinfo{author}{\bibfnamefont{P.}~\bibnamefont{Ruffieux}},
  \bibinfo{author}{\bibfnamefont{S.}~\bibnamefont{Wang}},
  \bibinfo{author}{\bibfnamefont{B.}~\bibnamefont{Yang}},
  \bibinfo{author}{\bibfnamefont{C.}~\bibnamefont{S{\'a}nchez-S{\'a}nchez}},
  \bibinfo{author}{\bibfnamefont{J.}~\bibnamefont{Liu}},
  \bibinfo{author}{\bibfnamefont{T.}~\bibnamefont{Dienel}},
  \bibinfo{author}{\bibfnamefont{L.}~\bibnamefont{Talirz}},
  \bibinfo{author}{\bibfnamefont{P.}~\bibnamefont{Shinde}},
  \bibinfo{author}{\bibfnamefont{C.~A.} \bibnamefont{Pignedoli}},
  \bibinfo{author}{\bibfnamefont{D.}~\bibnamefont{Passerone}},
  \bibnamefont{et~al.}, \bibinfo{journal}{Nature}
  \textbf{\bibinfo{volume}{531}}, \bibinfo{pages}{489} (\bibinfo{year}{2016}).

\bibitem[{\citenamefont{Su et~al.}(2020)\citenamefont{Su, Telychko, Song, and
  Lu}}]{su2020}
\bibinfo{author}{\bibfnamefont{J.}~\bibnamefont{Su}},
  \bibinfo{author}{\bibfnamefont{M.}~\bibnamefont{Telychko}},
  \bibinfo{author}{\bibfnamefont{S.}~\bibnamefont{Song}}, \bibnamefont{and}
  \bibinfo{author}{\bibfnamefont{J.}~\bibnamefont{Lu}},
  \bibinfo{journal}{Angewandte Chemie} \textbf{\bibinfo{volume}{132}},
  \bibinfo{pages}{7730} (\bibinfo{year}{2020}).

\bibitem[{\citenamefont{Pavli{\v{c}}ek
  et~al.}(2017)\citenamefont{Pavli{\v{c}}ek, Mistry, Majzik, Moll, Meyer, Fox,
  and Gross}}]{pavlivcek2017}
\bibinfo{author}{\bibfnamefont{N.}~\bibnamefont{Pavli{\v{c}}ek}},
  \bibinfo{author}{\bibfnamefont{A.}~\bibnamefont{Mistry}},
  \bibinfo{author}{\bibfnamefont{Z.}~\bibnamefont{Majzik}},
  \bibinfo{author}{\bibfnamefont{N.}~\bibnamefont{Moll}},
  \bibinfo{author}{\bibfnamefont{G.}~\bibnamefont{Meyer}},
  \bibinfo{author}{\bibfnamefont{D.~J.} \bibnamefont{Fox}}, \bibnamefont{and}
  \bibinfo{author}{\bibfnamefont{L.}~\bibnamefont{Gross}},
  \bibinfo{journal}{Nature Nanotechnology} \textbf{\bibinfo{volume}{12}},
  \bibinfo{pages}{308} (\bibinfo{year}{2017}).

\bibitem[{\citenamefont{Mishra et~al.}(2019)\citenamefont{Mishra, Beyer, Eimre,
  Liu, Berger, Groning, Pignedoli, M{\"u}llen, Fasel, Feng
  et~al.}}]{mishra2019b}
\bibinfo{author}{\bibfnamefont{S.}~\bibnamefont{Mishra}},
  \bibinfo{author}{\bibfnamefont{D.}~\bibnamefont{Beyer}},
  \bibinfo{author}{\bibfnamefont{K.}~\bibnamefont{Eimre}},
  \bibinfo{author}{\bibfnamefont{J.}~\bibnamefont{Liu}},
  \bibinfo{author}{\bibfnamefont{R.}~\bibnamefont{Berger}},
  \bibinfo{author}{\bibfnamefont{O.}~\bibnamefont{Groning}},
  \bibinfo{author}{\bibfnamefont{C.~A.} \bibnamefont{Pignedoli}},
  \bibinfo{author}{\bibfnamefont{K.}~\bibnamefont{M{\"u}llen}},
  \bibinfo{author}{\bibfnamefont{R.}~\bibnamefont{Fasel}},
  \bibinfo{author}{\bibfnamefont{X.}~\bibnamefont{Feng}}, \bibnamefont{et~al.},
  \bibinfo{journal}{Journal of the American Chemical Society}
  \textbf{\bibinfo{volume}{141}}, \bibinfo{pages}{10621}
  (\bibinfo{year}{2019}).

\bibitem[{\citenamefont{Su et~al.}(2019)\citenamefont{Su, Telychko, Hu, Macam,
  Mutombo, Zhang, Bao, Cheng, Huang, Qiu et~al.}}]{su2019}
\bibinfo{author}{\bibfnamefont{J.}~\bibnamefont{Su}},
  \bibinfo{author}{\bibfnamefont{M.}~\bibnamefont{Telychko}},
  \bibinfo{author}{\bibfnamefont{P.}~\bibnamefont{Hu}},
  \bibinfo{author}{\bibfnamefont{G.}~\bibnamefont{Macam}},
  \bibinfo{author}{\bibfnamefont{P.}~\bibnamefont{Mutombo}},
  \bibinfo{author}{\bibfnamefont{H.}~\bibnamefont{Zhang}},
  \bibinfo{author}{\bibfnamefont{Y.}~\bibnamefont{Bao}},
  \bibinfo{author}{\bibfnamefont{F.}~\bibnamefont{Cheng}},
  \bibinfo{author}{\bibfnamefont{Z.-Q.} \bibnamefont{Huang}},
  \bibinfo{author}{\bibfnamefont{Z.}~\bibnamefont{Qiu}}, \bibnamefont{et~al.},
  \bibinfo{journal}{Science advances} \textbf{\bibinfo{volume}{5}},
  \bibinfo{pages}{eaav7717} (\bibinfo{year}{2019}).

\bibitem[{\citenamefont{Mishra et~al.}(2021{\natexlab{a}})\citenamefont{Mishra,
  Xu, Eimre, Komber, Ma, Pignedoli, Fasel, Feng, and Ruffieux}}]{Mishra2021b}
\bibinfo{author}{\bibfnamefont{S.}~\bibnamefont{Mishra}},
  \bibinfo{author}{\bibfnamefont{K.}~\bibnamefont{Xu}},
  \bibinfo{author}{\bibfnamefont{K.}~\bibnamefont{Eimre}},
  \bibinfo{author}{\bibfnamefont{H.}~\bibnamefont{Komber}},
  \bibinfo{author}{\bibfnamefont{J.}~\bibnamefont{Ma}},
  \bibinfo{author}{\bibfnamefont{C.~A.} \bibnamefont{Pignedoli}},
  \bibinfo{author}{\bibfnamefont{R.}~\bibnamefont{Fasel}},
  \bibinfo{author}{\bibfnamefont{X.}~\bibnamefont{Feng}}, \bibnamefont{and}
  \bibinfo{author}{\bibfnamefont{P.}~\bibnamefont{Ruffieux}},
  \bibinfo{journal}{Nanoscale} \textbf{\bibinfo{volume}{13}},
  \bibinfo{pages}{1624} (\bibinfo{year}{2021}{\natexlab{a}}).

\bibitem[{\citenamefont{Mishra et~al.}(2020)\citenamefont{Mishra, Beyer, Eimre,
  Ortiz, Fern{\'a}ndez-Rossier, Berger, Gr{\"o}ning, Pignedoli, Fasel, Feng
  et~al.}}]{Mishra2020}
\bibinfo{author}{\bibfnamefont{S.}~\bibnamefont{Mishra}},
  \bibinfo{author}{\bibfnamefont{D.}~\bibnamefont{Beyer}},
  \bibinfo{author}{\bibfnamefont{K.}~\bibnamefont{Eimre}},
  \bibinfo{author}{\bibfnamefont{R.}~\bibnamefont{Ortiz}},
  \bibinfo{author}{\bibfnamefont{J.}~\bibnamefont{Fern{\'a}ndez-Rossier}},
  \bibinfo{author}{\bibfnamefont{R.}~\bibnamefont{Berger}},
  \bibinfo{author}{\bibfnamefont{O.}~\bibnamefont{Gr{\"o}ning}},
  \bibinfo{author}{\bibfnamefont{C.~A.} \bibnamefont{Pignedoli}},
  \bibinfo{author}{\bibfnamefont{R.}~\bibnamefont{Fasel}},
  \bibinfo{author}{\bibfnamefont{X.}~\bibnamefont{Feng}}, \bibnamefont{et~al.},
  \bibinfo{journal}{Angewandte Chemie International Edition}
  (\bibinfo{year}{2020}).

\bibitem[{\citenamefont{Mishra et~al.}(2021{\natexlab{b}})\citenamefont{Mishra,
  Catarina, Wu, Ortiz, Jacob, Eimre, Ma, Pignedoli, Feng, Ruffieux
  et~al.}}]{Mishra2021}
\bibinfo{author}{\bibfnamefont{S.}~\bibnamefont{Mishra}},
  \bibinfo{author}{\bibfnamefont{G.}~\bibnamefont{Catarina}},
  \bibinfo{author}{\bibfnamefont{F.}~\bibnamefont{Wu}},
  \bibinfo{author}{\bibfnamefont{R.}~\bibnamefont{Ortiz}},
  \bibinfo{author}{\bibfnamefont{D.}~\bibnamefont{Jacob}},
  \bibinfo{author}{\bibfnamefont{K.}~\bibnamefont{Eimre}},
  \bibinfo{author}{\bibfnamefont{J.}~\bibnamefont{Ma}},
  \bibinfo{author}{\bibfnamefont{C.~A.} \bibnamefont{Pignedoli}},
  \bibinfo{author}{\bibfnamefont{X.}~\bibnamefont{Feng}},
  \bibinfo{author}{\bibfnamefont{P.}~\bibnamefont{Ruffieux}},
  \bibnamefont{et~al.}, \bibinfo{journal}{Nature}
  \textbf{\bibinfo{volume}{598}}, \bibinfo{pages}{287}
  (\bibinfo{year}{2021}{\natexlab{b}}), ISSN \bibinfo{issn}{1476-4687}.

\bibitem[{\citenamefont{Hieulle et~al.}(2021)\citenamefont{Hieulle, Castro,
  Friedrich, Vegliante, Lara, Sanz, Rey, Corso, Frederiksen, Pascual
  et~al.}}]{hieulle2021}
\bibinfo{author}{\bibfnamefont{J.}~\bibnamefont{Hieulle}},
  \bibinfo{author}{\bibfnamefont{S.}~\bibnamefont{Castro}},
  \bibinfo{author}{\bibfnamefont{N.}~\bibnamefont{Friedrich}},
  \bibinfo{author}{\bibfnamefont{A.}~\bibnamefont{Vegliante}},
  \bibinfo{author}{\bibfnamefont{F.~R.} \bibnamefont{Lara}},
  \bibinfo{author}{\bibfnamefont{S.}~\bibnamefont{Sanz}},
  \bibinfo{author}{\bibfnamefont{D.}~\bibnamefont{Rey}},
  \bibinfo{author}{\bibfnamefont{M.}~\bibnamefont{Corso}},
  \bibinfo{author}{\bibfnamefont{T.}~\bibnamefont{Frederiksen}},
  \bibinfo{author}{\bibfnamefont{J.~I.} \bibnamefont{Pascual}},
  \bibnamefont{et~al.}, \bibinfo{journal}{Angewandte Chemie International
  Edition} \textbf{\bibinfo{volume}{60}}, \bibinfo{pages}{25224}
  (\bibinfo{year}{2021}).

\bibitem[{\citenamefont{Cheng et~al.}(2022)\citenamefont{Cheng, Xue, Li, Liu,
  Xiang, Ke, Yan, Wang, and Yu}}]{Cheng2022}
\bibinfo{author}{\bibfnamefont{S.}~\bibnamefont{Cheng}},
  \bibinfo{author}{\bibfnamefont{Z.}~\bibnamefont{Xue}},
  \bibinfo{author}{\bibfnamefont{C.}~\bibnamefont{Li}},
  \bibinfo{author}{\bibfnamefont{Y.}~\bibnamefont{Liu}},
  \bibinfo{author}{\bibfnamefont{L.}~\bibnamefont{Xiang}},
  \bibinfo{author}{\bibfnamefont{Y.}~\bibnamefont{Ke}},
  \bibinfo{author}{\bibfnamefont{K.}~\bibnamefont{Yan}},
  \bibinfo{author}{\bibfnamefont{S.}~\bibnamefont{Wang}}, \bibnamefont{and}
  \bibinfo{author}{\bibfnamefont{P.}~\bibnamefont{Yu}},
  \bibinfo{journal}{Nature Communications} \textbf{\bibinfo{volume}{13}},
  \bibinfo{pages}{1705} (\bibinfo{year}{2022}), ISSN \bibinfo{issn}{2041-1723}.

\bibitem[{\citenamefont{Steiner et~al.}(2017)\citenamefont{Steiner, Gebhardt,
  Ammon, Yang, Heidenreich, Hammer, G{\"o}rling, Kivala, and
  Maier}}]{steiner2017}
\bibinfo{author}{\bibfnamefont{C.}~\bibnamefont{Steiner}},
  \bibinfo{author}{\bibfnamefont{J.}~\bibnamefont{Gebhardt}},
  \bibinfo{author}{\bibfnamefont{M.}~\bibnamefont{Ammon}},
  \bibinfo{author}{\bibfnamefont{Z.}~\bibnamefont{Yang}},
  \bibinfo{author}{\bibfnamefont{A.}~\bibnamefont{Heidenreich}},
  \bibinfo{author}{\bibfnamefont{N.}~\bibnamefont{Hammer}},
  \bibinfo{author}{\bibfnamefont{A.}~\bibnamefont{G{\"o}rling}},
  \bibinfo{author}{\bibfnamefont{M.}~\bibnamefont{Kivala}}, \bibnamefont{and}
  \bibinfo{author}{\bibfnamefont{S.}~\bibnamefont{Maier}},
  \bibinfo{journal}{Nature communications} \textbf{\bibinfo{volume}{8}},
  \bibinfo{pages}{1} (\bibinfo{year}{2017}).

\bibitem[{\citenamefont{Moreno et~al.}(2018)\citenamefont{Moreno, Vilas-Varela,
  Kretz, Garcia-Lekue, Costache, Paradinas, Panighel, Ceballos, Valenzuela,
  Pe{\~n}a et~al.}}]{moreno2018}
\bibinfo{author}{\bibfnamefont{C.}~\bibnamefont{Moreno}},
  \bibinfo{author}{\bibfnamefont{M.}~\bibnamefont{Vilas-Varela}},
  \bibinfo{author}{\bibfnamefont{B.}~\bibnamefont{Kretz}},
  \bibinfo{author}{\bibfnamefont{A.}~\bibnamefont{Garcia-Lekue}},
  \bibinfo{author}{\bibfnamefont{M.~V.} \bibnamefont{Costache}},
  \bibinfo{author}{\bibfnamefont{M.}~\bibnamefont{Paradinas}},
  \bibinfo{author}{\bibfnamefont{M.}~\bibnamefont{Panighel}},
  \bibinfo{author}{\bibfnamefont{G.}~\bibnamefont{Ceballos}},
  \bibinfo{author}{\bibfnamefont{S.~O.} \bibnamefont{Valenzuela}},
  \bibinfo{author}{\bibfnamefont{D.}~\bibnamefont{Pe{\~n}a}},
  \bibnamefont{et~al.}, \bibinfo{journal}{Science}
  \textbf{\bibinfo{volume}{360}}, \bibinfo{pages}{199} (\bibinfo{year}{2018}).

\bibitem[{\citenamefont{Telychko et~al.}(2021)\citenamefont{Telychko, Li,
  Mutombo, Soler-Polo, Peng, Su, Song, Koh, Edmonds, Jel{\'\i}nek
  et~al.}}]{telychko2021}
\bibinfo{author}{\bibfnamefont{M.}~\bibnamefont{Telychko}},
  \bibinfo{author}{\bibfnamefont{G.}~\bibnamefont{Li}},
  \bibinfo{author}{\bibfnamefont{P.}~\bibnamefont{Mutombo}},
  \bibinfo{author}{\bibfnamefont{D.}~\bibnamefont{Soler-Polo}},
  \bibinfo{author}{\bibfnamefont{X.}~\bibnamefont{Peng}},
  \bibinfo{author}{\bibfnamefont{J.}~\bibnamefont{Su}},
  \bibinfo{author}{\bibfnamefont{S.}~\bibnamefont{Song}},
  \bibinfo{author}{\bibfnamefont{M.~J.} \bibnamefont{Koh}},
  \bibinfo{author}{\bibfnamefont{M.}~\bibnamefont{Edmonds}},
  \bibinfo{author}{\bibfnamefont{P.}~\bibnamefont{Jel{\'\i}nek}},
  \bibnamefont{et~al.}, \bibinfo{journal}{Science advances}
  \textbf{\bibinfo{volume}{7}}, \bibinfo{pages}{eabf0269}
  (\bibinfo{year}{2021}).

\bibitem[{\citenamefont{Ovchinnikov}(1978)}]{ovchinnikov78}
\bibinfo{author}{\bibfnamefont{A.~A.} \bibnamefont{Ovchinnikov}},
  \bibinfo{journal}{Theoretica Chimica Acta} \textbf{\bibinfo{volume}{47}},
  \bibinfo{pages}{297} (\bibinfo{year}{1978}).

\bibitem[{\citenamefont{Fern{\'a}ndez-Rossier and Palacios}(2007)}]{JFR07}
\bibinfo{author}{\bibfnamefont{J.}~\bibnamefont{Fern{\'a}ndez-Rossier}}
  \bibnamefont{and} \bibinfo{author}{\bibfnamefont{J.~J.}
  \bibnamefont{Palacios}}, \bibinfo{journal}{Physical Review Letters}
  \textbf{\bibinfo{volume}{99}}, \bibinfo{pages}{177204}
  (\bibinfo{year}{2007}).

\bibitem[{\citenamefont{Lado and Fern\'andez-Rossier}(2014)}]{Lado2014prl}
\bibinfo{author}{\bibfnamefont{J.~L.} \bibnamefont{Lado}} \bibnamefont{and}
  \bibinfo{author}{\bibfnamefont{J.}~\bibnamefont{Fern\'andez-Rossier}},
  \bibinfo{journal}{PRL} \textbf{\bibinfo{volume}{113}},
  \bibinfo{pages}{027203} (\bibinfo{year}{2014}).

\bibitem[{\citenamefont{Jacob et~al.}(2021)\citenamefont{Jacob, Ortiz, and
  Fern\'andez-Rossier}}]{jacob2021}
\bibinfo{author}{\bibfnamefont{D.}~\bibnamefont{Jacob}},
  \bibinfo{author}{\bibfnamefont{R.}~\bibnamefont{Ortiz}}, \bibnamefont{and}
  \bibinfo{author}{\bibfnamefont{J.}~\bibnamefont{Fern\'andez-Rossier}},
  \bibinfo{journal}{Phys. Rev. B} \textbf{\bibinfo{volume}{104}},
  \bibinfo{pages}{075404} (\bibinfo{year}{2021}).

\bibitem[{\citenamefont{Ortiz et~al.}(2019)\citenamefont{Ortiz, Boto,
  Garc{\'\i}a-Mart{\'\i}nez, Sancho-Garc{\'\i}a, Melle-Franco, and
  Fern{\'a}ndez-Rossier}}]{Ortiz19}
\bibinfo{author}{\bibfnamefont{R.}~\bibnamefont{Ortiz}},
  \bibinfo{author}{\bibfnamefont{R.~{\'A}.} \bibnamefont{Boto}},
  \bibinfo{author}{\bibfnamefont{N.}~\bibnamefont{Garc{\'\i}a-Mart{\'\i}nez}},
  \bibinfo{author}{\bibfnamefont{J.~C.} \bibnamefont{Sancho-Garc{\'\i}a}},
  \bibinfo{author}{\bibfnamefont{M.}~\bibnamefont{Melle-Franco}},
  \bibnamefont{and}
  \bibinfo{author}{\bibfnamefont{J.}~\bibnamefont{Fern{\'a}ndez-Rossier}},
  \bibinfo{journal}{Nano Lett.} \textbf{\bibinfo{volume}{19}},
  \bibinfo{pages}{5991} (\bibinfo{year}{2019}).

\bibitem[{\citenamefont{Wang et~al.}(2009)\citenamefont{Wang, Yazyev, Meng, and
  Kaxiras}}]{wang09}
\bibinfo{author}{\bibfnamefont{W.~L.} \bibnamefont{Wang}},
  \bibinfo{author}{\bibfnamefont{O.~V.} \bibnamefont{Yazyev}},
  \bibinfo{author}{\bibfnamefont{S.}~\bibnamefont{Meng}}, \bibnamefont{and}
  \bibinfo{author}{\bibfnamefont{E.}~\bibnamefont{Kaxiras}},
  \bibinfo{journal}{Physical review letters} \textbf{\bibinfo{volume}{102}},
  \bibinfo{pages}{157201} (\bibinfo{year}{2009}).

\bibitem[{\citenamefont{G\"u\ifmmode~\mbox{\c{c}}\else \c{c}\fi{}l\"u
  et~al.}(2009)\citenamefont{G\"u\ifmmode~\mbox{\c{c}}\else \c{c}\fi{}l\"u,
  Potasz, Voznyy, Korkusinski, and Hawrylak}}]{guclu09}
\bibinfo{author}{\bibfnamefont{A.~D.}
  \bibnamefont{G\"u\ifmmode~\mbox{\c{c}}\else \c{c}\fi{}l\"u}},
  \bibinfo{author}{\bibfnamefont{P.}~\bibnamefont{Potasz}},
  \bibinfo{author}{\bibfnamefont{O.}~\bibnamefont{Voznyy}},
  \bibinfo{author}{\bibfnamefont{M.}~\bibnamefont{Korkusinski}},
  \bibnamefont{and} \bibinfo{author}{\bibfnamefont{P.}~\bibnamefont{Hawrylak}},
  \bibinfo{journal}{Phys. Rev. Lett.} \textbf{\bibinfo{volume}{103}},
  \bibinfo{pages}{246805} (\bibinfo{year}{2009}).

\bibitem[{\citenamefont{Lieb}(1989)}]{Lieb89}
\bibinfo{author}{\bibfnamefont{E.~H.} \bibnamefont{Lieb}},
  \bibinfo{journal}{Physical Review Letters} \textbf{\bibinfo{volume}{62}},
  \bibinfo{pages}{1201} (\bibinfo{year}{1989}).

\bibitem[{\citenamefont{Sutherland}(1986)}]{Sutherland1986}
\bibinfo{author}{\bibfnamefont{B.}~\bibnamefont{Sutherland}},
  \bibinfo{journal}{Phys. Rev. B} \textbf{\bibinfo{volume}{34}},
  \bibinfo{pages}{5208} (\bibinfo{year}{1986}).

\bibitem[{\citenamefont{Giannozzi et~al.}(2009)\citenamefont{Giannozzi, Baroni,
  Bonini, Calandra, Car, Cavazzoni, Ceresoli, Chiarotti, Cococcioni, Dabo
  et~al.}}]{giannozzi2009quantum}
\bibinfo{author}{\bibfnamefont{P.}~\bibnamefont{Giannozzi}},
  \bibinfo{author}{\bibfnamefont{S.}~\bibnamefont{Baroni}},
  \bibinfo{author}{\bibfnamefont{N.}~\bibnamefont{Bonini}},
  \bibinfo{author}{\bibfnamefont{M.}~\bibnamefont{Calandra}},
  \bibinfo{author}{\bibfnamefont{R.}~\bibnamefont{Car}},
  \bibinfo{author}{\bibfnamefont{C.}~\bibnamefont{Cavazzoni}},
  \bibinfo{author}{\bibfnamefont{D.}~\bibnamefont{Ceresoli}},
  \bibinfo{author}{\bibfnamefont{G.~L.} \bibnamefont{Chiarotti}},
  \bibinfo{author}{\bibfnamefont{M.}~\bibnamefont{Cococcioni}},
  \bibinfo{author}{\bibfnamefont{I.}~\bibnamefont{Dabo}}, \bibnamefont{et~al.},
  \bibinfo{journal}{Journal of Physics: Condensed Matter}
  \textbf{\bibinfo{volume}{21}}, \bibinfo{pages}{395502}
  (\bibinfo{year}{2009}).

\bibitem[{\citenamefont{Perdew et~al.}(1996)\citenamefont{Perdew, Burke, and
  Ernzerhof}}]{perdew1996generalized}
\bibinfo{author}{\bibfnamefont{J.~P.} \bibnamefont{Perdew}},
  \bibinfo{author}{\bibfnamefont{K.}~\bibnamefont{Burke}}, \bibnamefont{and}
  \bibinfo{author}{\bibfnamefont{M.}~\bibnamefont{Ernzerhof}},
  \bibinfo{journal}{Physical review letters} \textbf{\bibinfo{volume}{77}},
  \bibinfo{pages}{3865} (\bibinfo{year}{1996}).

\bibitem[{\citenamefont{Zhou and Liu}(2020)}]{zhou20}
\bibinfo{author}{\bibfnamefont{Y.}~\bibnamefont{Zhou}} \bibnamefont{and}
  \bibinfo{author}{\bibfnamefont{F.}~\bibnamefont{Liu}}, \bibinfo{journal}{Nano
  Letters} \textbf{\bibinfo{volume}{21}}, \bibinfo{pages}{230}
  (\bibinfo{year}{2020}).

\bibitem[{\citenamefont{Sethi et~al.}(2021)\citenamefont{Sethi, Zhou, Zhu,
  Yang, and Liu}}]{sethi2021}
\bibinfo{author}{\bibfnamefont{G.}~\bibnamefont{Sethi}},
  \bibinfo{author}{\bibfnamefont{Y.}~\bibnamefont{Zhou}},
  \bibinfo{author}{\bibfnamefont{L.}~\bibnamefont{Zhu}},
  \bibinfo{author}{\bibfnamefont{L.}~\bibnamefont{Yang}}, \bibnamefont{and}
  \bibinfo{author}{\bibfnamefont{F.}~\bibnamefont{Liu}},
  \bibinfo{journal}{Physical Review Letters} \textbf{\bibinfo{volume}{126}},
  \bibinfo{pages}{196403} (\bibinfo{year}{2021}).

\bibitem[{\citenamefont{Wu et~al.}(2007)\citenamefont{Wu, Bergman, Balents, and
  Sarma}}]{wu2007}
\bibinfo{author}{\bibfnamefont{C.}~\bibnamefont{Wu}},
  \bibinfo{author}{\bibfnamefont{D.}~\bibnamefont{Bergman}},
  \bibinfo{author}{\bibfnamefont{L.}~\bibnamefont{Balents}}, \bibnamefont{and}
  \bibinfo{author}{\bibfnamefont{S.~D.} \bibnamefont{Sarma}},
  \bibinfo{journal}{Physical review letters} \textbf{\bibinfo{volume}{99}},
  \bibinfo{pages}{070401} (\bibinfo{year}{2007}).

\bibitem[{\citenamefont{Wu and Sarma}(2008)}]{wu2008}
\bibinfo{author}{\bibfnamefont{C.}~\bibnamefont{Wu}} \bibnamefont{and}
  \bibinfo{author}{\bibfnamefont{S.~D.} \bibnamefont{Sarma}},
  \bibinfo{journal}{Physical Review B} \textbf{\bibinfo{volume}{77}},
  \bibinfo{pages}{235107} (\bibinfo{year}{2008}).

\bibitem[{\citenamefont{Tamaki et~al.}(2020)\citenamefont{Tamaki, Kawakami, and
  Koshino}}]{tamaki2020}
\bibinfo{author}{\bibfnamefont{G.}~\bibnamefont{Tamaki}},
  \bibinfo{author}{\bibfnamefont{T.}~\bibnamefont{Kawakami}}, \bibnamefont{and}
  \bibinfo{author}{\bibfnamefont{M.}~\bibnamefont{Koshino}},
  \bibinfo{journal}{Physical Review B} \textbf{\bibinfo{volume}{101}},
  \bibinfo{pages}{205311} (\bibinfo{year}{2020}).

\bibitem[{\citenamefont{Tran et~al.}(2017)\citenamefont{Tran, Saint-Martin,
  Dollfus, and Volz}}]{Tran2017}
\bibinfo{author}{\bibfnamefont{V.-T.} \bibnamefont{Tran}},
  \bibinfo{author}{\bibfnamefont{J.}~\bibnamefont{Saint-Martin}},
  \bibinfo{author}{\bibfnamefont{P.}~\bibnamefont{Dollfus}}, \bibnamefont{and}
  \bibinfo{author}{\bibfnamefont{S.}~\bibnamefont{Volz}}, \bibinfo{journal}{AIP
  Advances} \textbf{\bibinfo{volume}{7}}, \bibinfo{pages}{075212}
  (\bibinfo{year}{2017}).

\bibitem[{\citenamefont{Mizoguchi et~al.}(2021)\citenamefont{Mizoguchi, Kuno,
  and Hatsugai}}]{Mizoguchi2021}
\bibinfo{author}{\bibfnamefont{T.}~\bibnamefont{Mizoguchi}},
  \bibinfo{author}{\bibfnamefont{Y.}~\bibnamefont{Kuno}}, \bibnamefont{and}
  \bibinfo{author}{\bibfnamefont{Y.}~\bibnamefont{Hatsugai}},
  \bibinfo{journal}{Phys. Rev. B} \textbf{\bibinfo{volume}{104}},
  \bibinfo{pages}{035161} (\bibinfo{year}{2021}).

\bibitem[{\citenamefont{Wang et~al.}(2022)\citenamefont{Wang,
  Berdonces-Layunta, Friedrich, Vilas-Varela, Calupitan, Pascual, Peña,
  Casanova, Corso, and de~Oteyza}}]{Wang2022}
\bibinfo{author}{\bibfnamefont{T.}~\bibnamefont{Wang}},
  \bibinfo{author}{\bibfnamefont{A.}~\bibnamefont{Berdonces-Layunta}},
  \bibinfo{author}{\bibfnamefont{N.}~\bibnamefont{Friedrich}},
  \bibinfo{author}{\bibfnamefont{M.}~\bibnamefont{Vilas-Varela}},
  \bibinfo{author}{\bibfnamefont{J.~P.} \bibnamefont{Calupitan}},
  \bibinfo{author}{\bibfnamefont{J.~I.} \bibnamefont{Pascual}},
  \bibinfo{author}{\bibfnamefont{D.}~\bibnamefont{Peña}},
  \bibinfo{author}{\bibfnamefont{D.}~\bibnamefont{Casanova}},
  \bibinfo{author}{\bibfnamefont{M.}~\bibnamefont{Corso}}, \bibnamefont{and}
  \bibinfo{author}{\bibfnamefont{D.~G.} \bibnamefont{de~Oteyza}},
  \bibinfo{journal}{Journal of the American Chemical Society}
  \textbf{\bibinfo{volume}{144}}, \bibinfo{pages}{4522} (\bibinfo{year}{2022}),
  \bibinfo{note}{pMID: 35254059},
  \eprint{https://doi.org/10.1021/jacs.1c12618},
  \urlprefix\url{https://doi.org/10.1021/jacs.1c12618}.

\bibitem[{\citenamefont{Sorella and Tosatti}(1992)}]{sorella92}
\bibinfo{author}{\bibfnamefont{S.}~\bibnamefont{Sorella}} \bibnamefont{and}
  \bibinfo{author}{\bibfnamefont{E.}~\bibnamefont{Tosatti}},
  \bibinfo{journal}{EPL (Europhysics Letters)} \textbf{\bibinfo{volume}{19}},
  \bibinfo{pages}{699} (\bibinfo{year}{1992}).

\bibitem[{\citenamefont{Jacob et~al.}()\citenamefont{Jacob, Ortiz, Catarina,
  and Fern\'andez-Rossier}}]{David22}
\bibinfo{author}{\bibfnamefont{D.}~\bibnamefont{Jacob}},
  \bibinfo{author}{\bibfnamefont{R.}~\bibnamefont{Ortiz}},
  \bibinfo{author}{\bibfnamefont{G.}~\bibnamefont{Catarina}}, \bibnamefont{and}
  \bibinfo{author}{\bibfnamefont{J.}~\bibnamefont{Fern\'andez-Rossier}},
  \emph{\bibinfo{title}{Theory of intermolecular exchange in nanographenes}}.

\bibitem[{\citenamefont{Gong et~al.}(2013)\citenamefont{Gong, Sheng, Motrunich,
  and Fisher}}]{gong2013}
\bibinfo{author}{\bibfnamefont{S.-S.} \bibnamefont{Gong}},
  \bibinfo{author}{\bibfnamefont{D.}~\bibnamefont{Sheng}},
  \bibinfo{author}{\bibfnamefont{O.~I.} \bibnamefont{Motrunich}},
  \bibnamefont{and} \bibinfo{author}{\bibfnamefont{M.~P.}
  \bibnamefont{Fisher}}, \bibinfo{journal}{Physical Review B}
  \textbf{\bibinfo{volume}{88}}, \bibinfo{pages}{165138}
  (\bibinfo{year}{2013}).

\bibitem[{\citenamefont{Gong et~al.}(2015)\citenamefont{Gong, Zhu, and
  Sheng}}]{gong2015}
\bibinfo{author}{\bibfnamefont{S.-S.} \bibnamefont{Gong}},
  \bibinfo{author}{\bibfnamefont{W.}~\bibnamefont{Zhu}}, \bibnamefont{and}
  \bibinfo{author}{\bibfnamefont{D.}~\bibnamefont{Sheng}},
  \bibinfo{journal}{Physical Review B} \textbf{\bibinfo{volume}{92}},
  \bibinfo{pages}{195110} (\bibinfo{year}{2015}).

\bibitem[{\citenamefont{Affleck et~al.}(1987)\citenamefont{Affleck, Kennedy,
  Lieb, and Tasaki}}]{Affleck1987}
\bibinfo{author}{\bibfnamefont{I.}~\bibnamefont{Affleck}},
  \bibinfo{author}{\bibfnamefont{T.}~\bibnamefont{Kennedy}},
  \bibinfo{author}{\bibfnamefont{E.~H.} \bibnamefont{Lieb}}, \bibnamefont{and}
  \bibinfo{author}{\bibfnamefont{H.}~\bibnamefont{Tasaki}},
  \bibinfo{journal}{Phys. Rev. Lett.} \textbf{\bibinfo{volume}{59}},
  \bibinfo{pages}{799} (\bibinfo{year}{1987}).

\bibitem[{\citenamefont{Wei et~al.}(2011)\citenamefont{Wei, Affleck, and
  Raussendorf}}]{Wei2011}
\bibinfo{author}{\bibfnamefont{T.-C.} \bibnamefont{Wei}},
  \bibinfo{author}{\bibfnamefont{I.}~\bibnamefont{Affleck}}, \bibnamefont{and}
  \bibinfo{author}{\bibfnamefont{R.}~\bibnamefont{Raussendorf}},
  \bibinfo{journal}{Phys. Rev. Lett.} \textbf{\bibinfo{volume}{106}},
  \bibinfo{pages}{070501} (\bibinfo{year}{2011}).

\bibitem[{\citenamefont{Garcia-Saez et~al.}(2013)\citenamefont{Garcia-Saez,
  Murg, and Wei}}]{Garcia-Saez2013}
\bibinfo{author}{\bibfnamefont{A.}~\bibnamefont{Garcia-Saez}},
  \bibinfo{author}{\bibfnamefont{V.}~\bibnamefont{Murg}}, \bibnamefont{and}
  \bibinfo{author}{\bibfnamefont{T.-C.} \bibnamefont{Wei}},
  \bibinfo{journal}{Phys. Rev. B} \textbf{\bibinfo{volume}{88}},
  \bibinfo{pages}{245118} (\bibinfo{year}{2013}).

\bibitem[{\citenamefont{Pomata and Wei}(2020)}]{pomata20}
\bibinfo{author}{\bibfnamefont{N.}~\bibnamefont{Pomata}} \bibnamefont{and}
  \bibinfo{author}{\bibfnamefont{T.-C.} \bibnamefont{Wei}},
  \bibinfo{journal}{Phys. Rev. Lett.} \textbf{\bibinfo{volume}{124}},
  \bibinfo{pages}{177203} (\bibinfo{year}{2020}),
  \urlprefix\url{https://link.aps.org/doi/10.1103/PhysRevLett.124.177203}.

\bibitem[{\citenamefont{Lemm et~al.}(2020)\citenamefont{Lemm, Sandvik, and
  Wang}}]{Lemm20}
\bibinfo{author}{\bibfnamefont{M.}~\bibnamefont{Lemm}},
  \bibinfo{author}{\bibfnamefont{A.~W.} \bibnamefont{Sandvik}},
  \bibnamefont{and} \bibinfo{author}{\bibfnamefont{L.}~\bibnamefont{Wang}},
  \bibinfo{journal}{Phys. Rev. Lett.} \textbf{\bibinfo{volume}{124}},
  \bibinfo{pages}{177204} (\bibinfo{year}{2020}),
  \urlprefix\url{https://link.aps.org/doi/10.1103/PhysRevLett.124.177204}.

\bibitem[{\citenamefont{Ganesh et~al.}(2011)\citenamefont{Ganesh, Sheng, Kim,
  and Paramekanti}}]{ganesh2011}
\bibinfo{author}{\bibfnamefont{R.}~\bibnamefont{Ganesh}},
  \bibinfo{author}{\bibfnamefont{D.}~\bibnamefont{Sheng}},
  \bibinfo{author}{\bibfnamefont{Y.-J.} \bibnamefont{Kim}}, \bibnamefont{and}
  \bibinfo{author}{\bibfnamefont{A.}~\bibnamefont{Paramekanti}},
  \bibinfo{journal}{Physical Review B} \textbf{\bibinfo{volume}{83}},
  \bibinfo{pages}{144414} (\bibinfo{year}{2011}).

\bibitem[{\citenamefont{Cao et~al.}(2018)\citenamefont{Cao, Fatemi, Fang,
  Watanabe, Taniguchi, Kaxiras, and Jarillo-Herrero}}]{cao2018b}
\bibinfo{author}{\bibfnamefont{Y.}~\bibnamefont{Cao}},
  \bibinfo{author}{\bibfnamefont{V.}~\bibnamefont{Fatemi}},
  \bibinfo{author}{\bibfnamefont{S.}~\bibnamefont{Fang}},
  \bibinfo{author}{\bibfnamefont{K.}~\bibnamefont{Watanabe}},
  \bibinfo{author}{\bibfnamefont{T.}~\bibnamefont{Taniguchi}},
  \bibinfo{author}{\bibfnamefont{E.}~\bibnamefont{Kaxiras}}, \bibnamefont{and}
  \bibinfo{author}{\bibfnamefont{P.}~\bibnamefont{Jarillo-Herrero}},
  \bibinfo{journal}{Nature} \textbf{\bibinfo{volume}{556}}, \bibinfo{pages}{43}
  (\bibinfo{year}{2018}).

\bibitem[{\citenamefont{Catarina and
  Fern{\'a}ndez-Rossier}(2022)}]{Catarina2022}
\bibinfo{author}{\bibfnamefont{G.}~\bibnamefont{Catarina}} \bibnamefont{and}
  \bibinfo{author}{\bibfnamefont{J.}~\bibnamefont{Fern{\'a}ndez-Rossier}},
  \bibinfo{journal}{Physical Review B} \textbf{\bibinfo{volume}{105}},
  \bibinfo{pages}{L081116} (\bibinfo{year}{2022}).

\end{thebibliography}

\end{document}